\documentclass[12pt]{article}
\usepackage{amsthm,amsmath,latexsym,amssymb,amsfonts,amscd}
\usepackage{graphics,lscape,fancyhdr,array,stmaryrd,euscript,wrapfig}
\usepackage{ifthen,empheq,slashed}
\usepackage{ifpdf}
\usepackage{verbatim,slashed}
\numberwithin{equation}{section}
\usepackage{hyperref,setspace}

\usepackage{mathrsfs}
\usepackage[numbers,sort&compress]{natbib}
\usepackage[nottoc]{tocbibind}

\newcommand{\pl}{\partial}

\newcommand{\be}{\begin{align}}
\newcommand{\ee}{\end{align}}

\newcommand{\aA}{{\ensuremath{\mathcal{A}}}}
\newcommand{\aB}{{\ensuremath{\mathcal{B}}}}

\newcommand{\fud}[2]{{}^{#1}{}_{#2}\,}
\newcommand{\fdu}[2]{{}_{#1}{}^{#2}\,}
\newcommand{\fudu}[3]{{}^{#1}{}_{#2}{}^{#3}\,}

\newcommand{\fdud}[3]{{}_{#1}{}^{#2}{}_{#3}\,}

\newcommand{\besubeqs}{\begin{subequations}}
\newcommand{\esubeqs}{\end{subequations}}

\usepackage{tensor}
\usepackage{todonotes}

\renewcommand{\bar}[1]{\overline{#1}}

\makeatletter
\newcommand{\subalign}[1]{%
  \vcenter{%
    \Let@ \restore@math@cr \default@tag
    \baselineskip\fontdimen10 \scriptfont\tw@
    \advance\baselineskip\fontdimen12 \scriptfont\tw@
    \lineskip\thr@@\fontdimen8 \scriptfont\thr@@
    \lineskiplimit\lineskip
    \ialign{\hfil$\m@th\scriptstyle##$&$\m@th\scriptstyle{}##$\crcr
      #1\crcr
    }%
  }
}
\makeatother

\begin{document}
\pagenumbering{gobble}
\hfill
\vskip 0.01\textheight
\begin{center}
{\Large\bfseries 
Minimal models of field theories: SDYM and SDGR}

\vspace{0.4cm}

\vskip 0.03\textheight
\renewcommand{\thefootnote}{\fnsymbol{footnote}}
Evgeny \textsc{Skvortsov}\footnote{Research Associate of the Fund for Scientific Research -- FNRS, Belgium}${}^{a,b}$ \& Richard Van Dongen${}^{a}$
\renewcommand{\thefootnote}{\arabic{footnote}}
\vskip 0.03\textheight

{\em ${}^{a}$ Service de Physique de l'Univers, Champs et Gravitation, \\ Universit\'e de Mons, 20 place du Parc, 7000 Mons, 
Belgium}\\
\vspace*{5pt}
{\em ${}^{b}$ Lebedev Institute of Physics, \\
Leninsky ave. 53, 119991 Moscow, Russia}\\

\end{center}

\vskip 0.02\textheight

\begin{abstract}
There exists a natural $L_\infty$-algebra or $Q$-manifold that can be associated to any (gauge) field theory. Perturbatively, it can be obtained by reducing the $L_\infty$-algebra behind the jet space BV-BRST formulation to its minimal model. We explicitly construct the minimal models of self-dual Yang-Mills and self-dual gravity theories, which also represents their equations of motion as Free Differential Algebras. The minimal model regains all relevant information about the field theory, e.g. actions, charges, anomalies, can be understood in terms of the corresponding $Q$-cohomology. 
\end{abstract}

\newpage
\tableofcontents
\newpage
\section{Introduction}
\pagenumbering{arabic}
\setcounter{page}{2}
Self-dual theories have a number of remarkable properties that make them very useful toy models in general and first order approximations to more complicated theories: 
(a) self-dual theories are closed subsectors of the corresponding complete theories; (b) as a result, all solutions of self-dual theories are solutions of the full ones; (c) all amplitudes of self-dual theories are also amplitudes of the full ones; (d) self-dual theories are integrable; (e) self-dual theories are finite and one-loop exact; (f) existence of a self-dual truncation allows one to rearrange the perturbation theory in a nontrivial way, e.g. to represent Yang-Mills theory as expansion over self-dual rather than flat backgrounds; (g) tools from twistor theory are very-well adapted to self-dual theories, see e.g. \cite{Penrose:1976js,Ward:1977ta,Atiyah:1978wi,Chalmers:1996rq,Mason:1991rf,Witten:2003nn,Berkovits:2004jj,Atiyah:2017erd}. In this letter we are interested in constructing $L_\infty$-algebras of the simplest self-dual theories: SDYM and SDGR, to uncover their algebraic structure.  

There is a hierarchy of $L_\infty$-algebras that originate from (quantum) field theories and string field theory, see e.g. \cite{Zwiebach:1992ie,Gaberdiel:1997ia,Kajiura:2003ax,Lada:1992wc,Alexandrov:1995kv,Barnich:2004cr,Hohm:2017pnh,Jurco:2018sby}. The simplest $L_\infty$-algebras emerge from a re-interpretation of the BV-BRST formalism: upon expanding the master action in ghosts and anti-fields one finds multilinear maps that obey $L_\infty$-relations. Another $L_\infty$-algebra emerges from the jet space version of the BV-BRST formulation of a given gauge theory \cite{Barnich:1994db,Barnich:1994mt,Barnich:2010sw,Grigoriev:2012xg,Grigoriev:2019ojp}. Such $L_\infty$ is especially useful when investigating various properties of this gauge theory systematically, e.g. classification of deformations of the action, or the question of possible anomalies \cite{Barnich:1994db,Barnich:1994mt}. Given an $L_\infty$-algebra one can consider various equivalent reductions. The smallest possible quasi-isomorphic algebra is the minimal model, which still captures all the relevant properties of the field theory. Another closely related algebraic structure is Free Differential Algebra \cite{Sullivan77}, which emerges as the sigma-model based on the minimal model. 

In this letter we construct the minimal models for self-dual Yang-Mills and self-dual gravity theories. As a starting point we take the Chalmers-Siegel action \cite{Chalmers:1996rq} for SDYM and the recently constructed action for SDGR with vanishing cosmological constant \cite{Krasnov:2021cva}, which is equivalent to other actions in the literature \cite{Siegel:1992wd,AbouZeid:2005dg}.  

Our general motivation stems from several possible applications, where we hope to understand from the algebraic, $L_\infty$, point of view: (i) integrability of self-dual theories; (ii) the double-copy relations, see \cite{Bern:2008qj,Bern:2010ue} and \cite{Campiglia:2021srh,Borsten:2021hua} for the recent results in this direction. Also, the results serve as a starting point for covariantization \cite{SDFDA2} of Chiral Higher Spin Gravity \cite{Metsaev:1991mt,Metsaev:1991nb,Ponomarev:2016lrm,Ponomarev:2017nrr,Skvortsov:2018jea,Skvortsov:2020wtf,Skvortsov:2020gpn}.

We begin with a short review of relation between $L_\infty$ and field theory and then proceed to SDYM and SDGR, respectively, with some technicalities left to appendices.

\section{Minimal models}
\label{sec:FDA}
As it was already sketched in the introduction, given any (gauge) field theory in the BV-BRST language it is natural to consider its jet space extension \cite{Brandt:1997iu,Brandt:1996mh,Barnich:2010sw,Grigoriev:2012xg,Grigoriev:2019ojp}, which is what is done when investigating the local BRST-cohomology \cite{Barnich:1994db,Barnich:1994mt}. The jet space extension leads to a rather big $L_\infty$-algebra, better say to a $Q$-manifold provided global issues are taken into account. Various $Q$-cohomology groups correspond to all physically relevant quantities, e.g. deformations/interactions, anomalies, charges, etc., see e.g. \cite{Barnich:1994db,Barnich:1994mt}. For every $L_\infty$-algebra there always exists a (usually much smaller) $L_\infty$-algebra, known as the minimal model, see e.g. \cite{Huebschmann,Grigoriev:2019ojp,Grigoriev:2020lzu}, that contains the same information --- it is said to be quasi-isomorphic.\footnote{There is also another, 'quantum', minimal model \cite{Arvanitakis:2019ald} --- the $L_\infty$-algebra given by 1PI correlation functions. }  Some care is needed to prove the same statement for field theories \cite{Barnich:2009jy,Grigoriev:2019ojp}, where relevant $L_\infty$-algebras are necessarily infinite-dimensional. Minimal models were first introduced by Sullivan \cite{Sullivan77} in the context of differential graded algebras to study rational homotopy theory. We construct such minimal models for SDYM and SDGR.

Given any non-negatively graded supermanifold $\mathcal{N}$ equipped with a homological vector field $Q$, $QQ=0$, e.g. given by the minimal model, one can write down a sigma-model \cite{Barnich:2010sw}:
\begin{align*}
    d \Phi&= Q(\Phi)  \,,
\end{align*}
where $\Phi\equiv \Phi^\aA$ are maps $\Pi T\mathcal{M} \rightarrow \mathcal{N}$ from the exterior algebra of differential forms on a space-time manifold $\mathcal{M}$ to $\mathcal{N}$. Together with natural gauge symmetries the sigma-model is equivalent to the classical equations of motion of the initial field theory \cite{Barnich:2010sw,Grigoriev:2012xg,Grigoriev:2019ojp}, thereby having the form of a Free Differential Algebra, see \cite{Sullivan77} for exact definitions.\footnote{FDA was introduced by Sullivan and applied to problems in topology. Later, FDA's sneaked into physics in the context of supersymmetry and supergravity \cite{vanNieuwenhuizen:1982zf,DAuria:1980cmy} and, even later, applied to construct formally consistent deformations of the FDA for free higher spin fields \cite{Vasiliev:1988sa}. } In the paper we adopt a more pragmatic point of view on minimal models: we seek for the classical equations of motion as an FDA \cite{Vasiliev:1988sa}. If $\Phi=\{ \Phi^\aA\}$ are coordinates on $\mathcal{N}$, then $Q=Q^\aA \pl/\pl \Phi^\aA$ and 
\begin{align*}
    Q^2&=0 &&\Longleftrightarrow && Q^\aB \tfrac{\pl}{\pl \Phi^\aB} Q^\aA=0\,.
\end{align*}
This condition is equivalent to the Frobenius integrability of the field equations, i.e. the equations are formally consistent. The $L_\infty$-relations emerge by Taylor expanding $QQ=0$ at a stationary point of $Q$ \cite{Alexandrov:1995kv}. By abuse of notation we always denote coordinates on $\mathcal{N}$ and the corresponding fields by the same symbols. For a large class of field theories $\mathcal{N}$ has coordinates of degree-one and degree-zero to be associated with gauge connection(s) $A$ and with some matter-like zero-forms $L$. The simplest FDA with this data reads  
\begin{align*}
    dA&=\tfrac12[A,A]\,,& dL&=\rho(A)L\,,
\end{align*}
and is equivalent to $A$ taking values in some Lie algebra and to $L$ taking values in its module $\rho$. We consider these equations free. In particular, it is easy to solve them locally in the pure gauge form, e.g. $A=g^{-1} dg$. The most general deformation of the free equations here-above that is consistent with the form-degree counting reads
\begin{align*}
     dA&=l_2(A,A)+l_3(A,A,L)+l_4(A,A,L,L)+\ldots=F_A(A;L)\,,\\
     dL&=l_2(A,L)+l_3(A,L,L)+\ldots=F_L(A;L)\,.     
\end{align*}
Our strategy for each of the cases, SDYM and SDGR, is to start off with an action, rewrite the variational equations of motion in the 'almost' FDA form, where 'almost' means that at each step the equations/$Q$-structure will require new fields/coordinates on $\mathcal{N}$ be introduced. At the end of the day we find the complete $Q$. Interacting field theories are defined modulo admissible field redefinitions (those that do not change the $S$-matrix). We found a field frame where no structure maps higher than $l_3(\bullet,\bullet,\bullet)$ are needed for SDYM and SDGR, which also fixes all field redefinitions.\footnote{This is a key difference with respect to \cite{Vasiliev:1988sa}, where locality and field redefinitions are not taken into account \cite{Boulanger:2015ova,Skvortsov:2015lja}, which results in a general ansatz for interactions rather than a concrete theory.  }

\section{SDYM}
\label{sec:SDYM}

\subsection{Action, initial data}

The theory can be formulated  \cite{Chalmers:1996rq}  with two fields:\footnote{We use almost exclusively the two-component spinor language, which is well-suited for $4d$-theories. A short compendium can be found in Appendix \ref{app:notation}. A classical source is \cite{penroserindler}. The most important fact about our notation is that symmetric or to be symmetrized indices can be denoted by the same letter. Also, $A(k)\equiv A_1...A_k$.  } the usual one-form gauge potential $A\equiv A_\mu \, dx^\mu \equiv A_\mu^a \, dx^\mu\, T_a$ and a zero-form $\Psi^{AB}\equiv\Psi^{BA}$, $\Psi^{AB}\equiv \Psi^{AB; a}\, T_a$. Here $T_a$ are generators of some Lie algebra with a non-degenerate invariant bilinear form. We usually suppress form indices and $dx$'s, as well as the Lie algebra indices. In practice, it is convenient to think of generators $T_a$ as of taking values in some matrix algebra and assume $A$ and $\Psi^{AB}$ to take values in $\mathrm{Mat}_N$, with matrix indices again suppressed. The action reads\footnote{Here, see also appendix \ref{app:notation}, $H^{AB}\equiv H^{BA}$, $H^{A'B'}\equiv H^{B'A'}$ is the basis of self-dual two-forms, $H^{AB}\equiv e\fud{A}{C'}\wedge e^{BC'}$, \textit{idem.} for $H^{A'B'}$. Vierbein one-form is $e^{AA'}$. }
\begin{align*}
    S_{SDYM}&=\mathrm{tr} \int \Psi^{A'B'} \wedge H_{A'B'} \wedge F \,,
\end{align*}
where $F=dA-A\wedge A$ and we prefer to omit $\wedge$-symbol. The equations of motion imply
\begin{align} \label{SDYMeq}
    F_{A'B'}&=0 \,, & D\fud{A}{B'}\Psi^{A'B'}&=0 \,,
\end{align}
where $D\equiv dx^{AA'}\,D_{AA'}\equiv \nabla- [A,\bullet]$ is the gauge and Lorentz covariant derivative. We also used the decomposition of $F$ into (anti)self-dual parts
\begin{equation*}
    F=H^{BB}F_{BB}+H^{B'B'}F_{B'B'} \,.
\end{equation*}
We can rewrite the variational equations as
\begin{align}\label{F&Psi}
    dA-AA&=H^{BB}F_{BB}\,, & D \Psi^{A'B'}&= e_{CC'}\Psi^{C,A'B'C'} \,,
\end{align}
which is the starting point for constructing the corresponding $L_\infty$-algebra. The first equation simply states that $F_{A'B'}=0$ and, hence, connection $A$ is self-dual. Therefore, only the self-dual part may not be trivial and it is parameterized by $F^{AB}$. A simple consequence is the Bianchi identity for $F^{AB}$. In the second equation we introduced a field $\Psi^{A,A'B'C'}$ that parameterizes the first derivative of $\Psi$ that is consistent with \eqref{SDYMeq}, i.e. it corresponds to a coordinate on the on-shell jet of $\Psi^{A'B'}$.\footnote{Equations of motion for free fields of arbitrary spin can be recast into the FDA form \cite{Vasiliev:1986td}. The on-shell jet is very easy to describe in spinorial language \cite{penroserindler}. } 

The problem is, therefore, to find a completion of \eqref{F&Psi}, which requires an infinite set of coordinates on $\mathcal{N}$ and $Q$ defined on them in such a way that $QQ=0$. The first few terms of $Q$ and $\mathcal{N}$ are already clear from \eqref{F&Psi}. The on-shell jet space is also well-known \cite{penroserindler}. It is the same as for the free theory where we turned off non-Abelian Yang-Mills groups that result in non-linearities. That the coordinates on $\mathcal{N}$ are the same for the free and interacting theories is due to the requirement for them to have the same number of local degrees of freedom. 

\paragraph{Coordinates, on-shell jet.} The coordinates on $\mathcal{N}$ are: degree-one $A$; degree-zero $F^{A(k+2),A'(k)}$ and $\Psi^{A(k),A'(k+2)}$, $k=0,1,2,...$. The free equations, i.e. (self-dual) Maxwell equations on Minkowski space, can be written as  \cite{Vasiliev:1986td}
\begin{align}\label{eqMaxwA}
    dA&=H^{BB}F_{BB} +\epsilon H^{B'B'} \Psi_{B'B'} \,,
\end{align}
which just defines $F^{AB}$ and $\Psi^{A'B'}$ as (anti)-self-dual components of $dA$. The Bianchi identities imply 
\begin{align}\label{eqMaxwB}
    dF^{A(k+2),A'(k)}=e_{BB'}F^{A(k+2)B,A'(k)B'}\,,
\end{align}
and a similar chain of equations for the field $\Psi$
\begin{align}\label{eqMaxwC}
    d\Psi^{A(k),A'(k+2)}=e_{CC'}\Psi^{A(k)C,A'(k+2)C'}\,.
\end{align}
The system \eqref{eqMaxwA}, \eqref{eqMaxwB}, \eqref{eqMaxwC} is equivalent to Maxwell equations, i.e. no self-dual truncation has yet been taken. The free SDYM equations are obtained by setting $\epsilon=0$ in \eqref{eqMaxwA}, while no other modifications are needed. What erasing $\Psi^{A'B'}$ from \eqref{eqMaxwA} does is that it makes the anti-selfdual part of $dA$ vanish. The $\Psi$-subsystem \eqref{eqMaxwC} decouples and describes the second degree of freedom (say, helicity $-1$). The first equations in \eqref{eqMaxwB} and \eqref{eqMaxwC} are equivalent to the well-known \cite{Penrose:1965am}
\begin{align*}
    D\fdu{A}{B'}F^{AB}&=0 \,, & D\fud{A}{B'}\Psi^{A'B'}&=0\,,
\end{align*}
and describe helicity $+1$ and $-1$ degrees of freedom. Subsystems \eqref{eqMaxwB} and \eqref{eqMaxwC} are closed and identical to each other (upon swapping primed and unprimed indices). What makes them different is that only the physical degree of freedom carried by $F$ gets embedded into $A$ once we set $\epsilon=0$. There is no change in the number of physical degrees of freedom in the $\epsilon=0$ limit.  

\paragraph{General form.} In order to have a genuine FDA we should incorporate the background gravitational fields: vierbein $e^{AA'}$ and the (anti)-self-dual components $\omega^{AB}$, $\omega^{A'B'}$ of spin-connection. Finally, we have
\begin{align*}
    \mathcal{N}&: && 
    \begin{aligned}
        1&: e^{AA'}\,, \omega^{AB}\,, \omega^{A'B'} \,, A \,,\\
        0&: F^{A(k+2),A'(k)}\,, \Psi^{A(k),A'(k+2)}\,, k=0,1,2,...
    \end{aligned}
\end{align*}
We will prove below that the complete $L_\infty$-algebra of SDYM can be cast into the following simple form:
\besubeqs
\begin{align*}
    d e^{AA'}&= \omega\fud{A}{B}\wedge
    e^{BA'}+\omega\fud{A'}{B'}\wedge e^{A B'} \,,\\  
    d\omega^{AB}&= \omega\fud{A}{C}\wedge \omega^{BC} \,,\\
    d\omega^{A'B'}&= \omega\fud{A'}{C'}\wedge \omega^{B'C'} \,,\\
    dA&= AA + H_{BB}F^{BB} \,,\\
    dF&= l_2(\omega,F)+l_2(A,F)+l_2(e,F)+l_3(e,F,F) \,,\\
    d\Psi&= l_2(\omega,\Psi)+l_2(A,\Psi)+l_2(e,\Psi) +l_3(e,F,\Psi) \,.
\end{align*}
\esubeqs
Some of the maps above are self-evident, e.g. $l_2(\omega,\bullet)$ and $l_2(A,\bullet)$ are parts of the usual Lorentz and gauge covariant derivatives. $H_{BB}F^{BB}$ is a specific tri-linear map $l_3(e,e,F)$. 
Introducing the standard Lorentz covariant derivative $\nabla$ and appending it with the gauge part $[A,\bullet]$ we define $D=\nabla -[A,\bullet]$. The equations reduce to
\besubeqs
\begin{align*}
    \nabla e^{AA'}&= 0 \,, & \nabla^2&=0 \,,\\
    dA&= AA + H_{BB}F^{BB} \,, \\
    DF&= l_2(e,F)+l_3(e,F,F) \,, \\
    D\Psi&= l_2(e,\Psi)+l_3(e,F,\Psi) \,.
\end{align*}
\esubeqs
The first line is equivalent to living in Minkowski space. Covariant derivative $D$ allows us to absorb $l_2(A,F)=[A,F]$ and $l_2(A,\Psi)=[A,\Psi]$. The $L_\infty$-structure relations are equivalent to (i) $e$, $\omega$ being a flat connection of Poincare algebra; (ii) a bit more nontrivial $L_\infty$-relations that follow from
\begin{align*}
    D^2 F +l_2(e, DF) +l_3(e,DF,F)+l_3(e,F,DF)&\equiv0 \,,
    \\
    D^2\Psi+l_2(e,D\Psi)+l_3(e,DF,\Psi)+l_3(e,F,D\Psi)&\equiv0
\end{align*}
and decompose into
{\allowdisplaybreaks
\besubeqs \label{SDYMstasheff}
\begin{align} 
    l_2(e, l_2(e,F))&\equiv0 \label{SDYMstasheffF1} \,, \\
    -[H_{BB}F^{BB}, F] +l_2(e, l_3(e,F,F)) +l_3(e,l_2(e,F),F)+l_3(e,F,l_2(e,F))&\equiv0 \label{SDYMstasheffF2} \,, \\
    l_3(e,l_3(e,F,F),F)+l_3(e,F,l_3(e,F,F))&\equiv0 \label{SDYMstasheffF3} \,, \\
    l_2(e,l_2(e,\Psi))&\equiv0 \label{SDYMstafhessPsi1} \,, \\
    -[H_{BB}F^{BB},\Psi]+l_2(e,l_3(e,F,\Psi))+l_3(e,l_2(e,F),\Psi)+l_3(e,F,l_2(e,\Psi))&\equiv0 \label{SDYMstafhessPsi2} \,,\\
    l_3(e,l_3(e,F,F),\Psi)+l_3(e,F,l_3(e,F,\Psi))&\equiv0 \label{SDYMstafhessPsi3}\,.
\end{align}
\esubeqs}
The first and the fourth relations are guaranteed by the free equations of motion. 

\subsection{FDA, flat space}
\label{sec:flatSDYM}
\paragraph{Appetizer.} Firstly, let us explain why a non-linear completion of \eqref{eqMaxwA}, \eqref{eqMaxwB}, \eqref{eqMaxwC} is necessary. The root of the nonlinear completion is in the fact that $D^2\neq0$ and, for a 
field $\chi$ in representation $\rho$ of the Yang-Mills algebra we find $D^2\chi=-\rho(F)\chi$. In the adjoint representation, one gets (matrix/Lie algebra indices are implicit)
\begin{align*}
	D^2\chi=-[F,\chi]=-H_{BB}[F^{BB},\chi] \,.
\end{align*}
Therefore, the Bianchi identity for the first equation in the $F$-subsystem
\begin{align*}
	D F^{AA}=e_{BB'}\wedge F^{AAB,B'}
\end{align*}
leads to 
\begin{align} \label{nablasquaredF}
	D^2 F^{AA}=-H_{BB}[F^{BB},F^{AA}]=-e_{BB'}\wedge D F^{AAB,B'}\,.
\end{align}
The aim is to use the above equation to obtain $D F^{AAA,A'}$. Matching the indices and imposing that $D F^{AAA,A'}$ is a one-form, one may take the ansatz
\begin{align*}
    \begin{aligned}
	   D F^{AAA,A'}&=e_{CC'}F^{AAAC,A'C'}+\alpha e\fdu{C}{A'}[F^{AC},F^{AA}]+\beta e\fdu{C}{A'}[F^{AA},F^{AC}]\\
	   &+\gamma e^{AA'}[F\fud{A}{C},F^{AC}] \,.
	 \end{aligned}
\end{align*}
The Fierz identity \eqref{Fierz} and the anti-symmetry of the commutator reduce this to
\begin{equation*}
    D F^{AAA,A'}=e_{CC'}F^{AAAC,A'C'}+\alpha_{00} e\fdu{C}{A'}[F^{AC},F^{AA}]\,,
\end{equation*}
where the label on $\alpha_{00}$ was added for future convenience. Upon contraction with $e_{BB'}$ this yields\footnote{The first term can be rewritten as $e_{BB'}\wedge e_{CC'}F^{AABC,B'C'}=\tfrac{1}{2}(H_{BC}\epsilon_{B'C'}+\epsilon_{BC}H_{B'C'})F^{AABC,B'C'}$  and vanishes as the contracted indices are symmetrized in $F^{AABC,B'C'}$ and anti-symmetrized in the $\epsilon$'s. The last term must be zero, because $e_{BB'}\wedge e\fdu{C}{B'}=\tfrac{1}{2}H_{BC}$ is symmetric in $B,C$, whereas the commutator is anti-symmetric. }
\begin{align} \label{e-nablaF3}
    \begin{aligned}
	    e_{BB'}\wedge D F^{AAB,B'}&=0+\tfrac{1}{3}\alpha_{00} e_{BB'}\wedge e\fdu{C}{B'}[F^{BC},F^{AA}]+\tfrac{2}{3}\alpha_{00}e_{BB'}\wedge e\fdu{C}{B'}[F^{AC},F^{AB}]\\
	    &=\tfrac{1}{3}\alpha_{00} H_{BB}[F^{BB},F^{AA}] \,,
    \end{aligned}
\end{align}
where \eqref{e-wedge-e} was used. Comparing this to \eqref{nablasquaredF}, one obtains the solution 
\begin{align} \label{DF3}
    D F^{AAA,A'}=e_{CC'}F^{AAAC,A'C'}+3e\fdu{C}{A'}[F^{AC},F^{AA}]\,,
\end{align}
where the first term on the r.h.s. is there since the free equations. Similarly, taking the covariant derivative of the above result yields another consistency equation. Following the same steps as before, one finds
\begin{align*}
    \begin{aligned}
	D^2 F^{AAA,A'}&=-H_{BB}[F^{BB},F^{AAA,A'}]=-e_{BB'}\wedge D F^{AAAB,A'B'}\\
	&-3e\fdu{C}{A'}\wedge[D F^{AC},F^{AA}]-3e\fdu{C}{A'}[F^{AC},D F^{AA}]\,,
	\end{aligned}
\end{align*}
which results in
\begin{align} \label{e-DF4}
    \begin{aligned}
    e_{BB'}\wedge D F^{AAAB,A'B'}&=H_{BB}[F^{BB},F^{AAA,A'}]-\tfrac{3}{2}H_{BB}[F^{AA},F^{ABB,A'}]
    \\
    &+\tfrac{3}{2}H_{BB}[F^{AB},F^{AAB,B'}]+\tfrac{3}{2}H\fdu{B'}{A'}[F^{AB},F\fudu{AA}{B}{,B'}]\,.
    \end{aligned}
\end{align}
The minimal ansatz for $DF^{AAAA,A'A'}$ reads
\begin{align*}
    \begin{aligned}
    D F^{AAAA,A'A'}&=e_{CC'}F^{AAAAC,A'A'C'}+\alpha_{02} e\fdu{C}{A'}[F^{AC},F^{AAA,A'}]\\
    &+\alpha_{12} e\fdu{C}{A'}[F^{AAC,A'},F^{AA}]\,.
    \end{aligned}
\end{align*}
We contract this with $e_{BB'}$ to find
\begin{align}
    \begin{aligned}
        e_{BB'}\wedge DF^{AAAB,A'B'} &= \tfrac{3\alpha_{02}}{16}H_{BB}[F^{BB},F^{AAA,A'}]+(\tfrac{9\alpha_{02}}{16}+\tfrac{3\alpha_{12}}{8})H_{BB}[F^{AB},F^{AAB,A'}] \\
        &+\tfrac{3\alpha_{12}}{8}H_{BB}[F^{AA},F^{ABB,A'}] + (\tfrac{3\alpha_{02}}{18}-\tfrac{\alpha_{12}}{8})H\fdu{B'}{A'}[F^{AB},F\fudu{AA}{B}{,B'}] \,.
    \end{aligned}
\end{align}
We compare this to \eqref{e-DF4} to obtain the result
\begin{align*}
    \begin{aligned}
        D F^{AAAA,A'A'}&=e_{CC'}F^{AAAAC,A'A'C'}+\tfrac{16}{3}e\fdu{C}{A'}[F^{AC},F^{AAA,A'}]\\
        &+4e\fdu{C}{A'}[F^{AAC,A'},F^{AA}]\,.
    \end{aligned}
\end{align*}
The procedure presented above is nothing more than the practical realisation of solving the $L_\infty$-relation \eqref{SDYMstasheffF2}. This procedure will be generalized next. 

\paragraph{Main course, $\boldsymbol{F}$-sector.} By looking at the first few equations in the system it is easy to come up with an ansatz:
\begin{align}\label{spin1ansatz}
\begin{aligned}
        D F_{A(k+2),A'(k)}&=e^{BB'}F_{A(k+2)B,A'(k)B'}\\
        &+\sum_{n=0}^{k-1}\alpha_{nk}e\fud{B}{A'}[F_{A(n+1)B,A'(n)},F_{A(k-n+1),A'(k-n-1)}] \,,   
\end{aligned}
\end{align}
for any $k\geq 0$. This ansatz makes use of the fact that $D F_{A(k+2),A'(k)}$ should be a one-form, which requires the presence of $e^{BB'}$ and it matches the number of (un)-primed indices. In any non-linear theory there is always a freedom to perform field redefinitions. We have also fixed the redefinitions by requiring that there are no index contractions between $F$ in $[F,F]$. Terms with contracted indices can easily be introduced by field-redefinitions. Our ansatz contains only the terms that are necessary to ensure consistency and, thereby, is the minimal one.  

Taking the covariant derivative of the ansatz yields
\begin{align} \label{DsquaredF}
    \begin{aligned}
        D^2F_{A(k+2),A'(k)}=&-H^{BB}[F_{BB},F_{A(k+2),A'(k)}]=-e^{BB'}\wedge DF_{A(k+2)B,A'(k)B'}\\
        &-e\fud{B}{A'}\wedge\sum_{n=0}^{k-1}\alpha_{nk}[DF_{A(n+1)B,A'(n)},F_{A(k-n+1),A'(k-n-1)}] \\
        &-e\fud{B}{A'}\wedge\sum_{n=0}^{k-1}\alpha_{nk}[F_{A(n+1)B,A'(n)},DF_{A(k-n+1),A'(k-n-1)}] \,.
    \end{aligned}
\end{align}
and considering only terms quadratic in $F$ gives\footnote{In the third term we have made the anti-symmetry of the commutator explicit by writing $[X,Y]=\tfrac{1}{2}([X,Y]-[Y,X])$ and renaming the dummy indices accordingly. This automatically gets rid of terms that vanish because of symmetry reasons, like the last term in the middle expression of \eqref{e-nablaF3}. As the summation now runs up to $n=k$, the coefficient $\alpha_{kk}$ shows up, so we set $\alpha_{kk}=0$ by hand since it was not present in the ansatz.}
\begin{align} \label{e-nabla-F1}
    \begin{aligned}
    e^{BB'}\wedge D F_{A(k+2)B,A'(k)B'}&=H^{BB}[F_{BB},F_{A(k+2),A'(k)}]\\
    &-\tfrac{1}{2}H^{BB}\sum_{n=0}^{k-1}\alpha_{nk}[F_{A(n+1)BB,A'(n+1)},F_{A(k-n+1),A'(k-n-1)}]\\
    &-\tfrac{1}{4}H^{BB}\sum_{n=0}^{k}(\alpha_{nk}-\alpha_{(k-n)k})[F_{A(n+1)B,A'(n)},F_{A(k-n+1)B,A'(k-n)}]\\
    &+\tfrac{1}{2}H\fud{B'}{A'}\sum_{n=0}^{k-1}\alpha_{nk}[F\fdud{A(n+1)}{B}{,A'(n)},F_{A(k-n+1)B,A'(k-n-1)B'}] \,,
    \end{aligned}
\end{align}
where terms cubic in $F$ are ignored for now. Alternatively, we contract $e^{BB'}$ with $D F_{A(k+3),A'(k+1)}$ to obtain
\allowdisplaybreaks{
\begin{align*}
    &e^{BB'}\wedge D F_{A(k+2)B,A'(k)B'}=-\tfrac{1}{2}H^{BB}\alpha_{0(k+1)}\tfrac{k+2}{(k+3)(k+1)}[F_{BB},F_{A(k+2),A'(k)}]\\
    &-\tfrac{1}{2}H^{BB}\sum_{n=0}^{k-1}\alpha_{(n+1)(k+1)}\tfrac{(n+2)(k+2)}{(k+3)(k+1)}[F_{A(n+1)BB,A'(n+1)},F_{A(k-n+1),A'(k-n-1)}]\\
    &-\tfrac{1}{4}H^{BB}\sum_{n=0}^{k}(\alpha_{n(k+1)}\tfrac{(k-n+2)(k+2)}{(k+3)(k+1)}-\alpha_{(k-n)(k+1)})\tfrac{(n+2)(k+2)}{(k+3)(k+1)})[F_{A(n+1)B,A'(n)},F_{A(k-n+1)B,A'(k-n)}]\\
    &+\tfrac{1}{2}H\fud{B'}{A'}\sum_{n=0}^{k}(\alpha_{(k-n)(k+1)}\tfrac{(n+2)(k-n)}{(k+3)(k+1)}+\alpha_{n(k+1)}\tfrac{(k-n+2)(k-n)}{(k+3)(k+1)})[F\fdud{A(n+1)}{B}{,A'(n)},F_{A(k-n+1)B,A'(k-n-1)B'}] \,.
\end{align*}}
Comparing this with \eqref{e-nabla-F1} results in the following system of recurrence relations:
\besubeqs
    \begin{align*}
        0&=\alpha_{0k}+\tfrac{2k(k+2)}{k+1}\,,\\
        0&=\alpha_{(n+1)(k+1)}\tfrac{(n+2)(k+2)}{(k+3)(k+1)}-\alpha_{nk}\,,\\
        0&=\alpha_{n(k+1)}\tfrac{(k-n+2)(k+2)}{(k+3)(k+1)}-\alpha_{(k-n)(k+1)}\tfrac{(n+2)(k+2)}{(k+3)(k+1)}-\alpha_{nk}+\alpha_{(k-n)k}\,,\\
        0&=\alpha_{(k-n)(k+1)}\tfrac{(n+2)(k-n)}{(k+3)(k+1)}+\alpha_{n(k+1)}\tfrac{(k-n+2)(k-n)}{(k+3)(k+1)}-\alpha_{nk}\,.
    \end{align*}
\esubeqs
This system is over-determined, but, nevertheless,  is solved by
\begin{align*}
    \alpha_{nk}=-\tfrac{2}{(n+1)!}\tfrac{(k+2)!}{(k-n-1)!(k-n+1)(k+1)}\,.
\end{align*}
The full solution reads
\begin{align} \label{spin1sol}\boxed{
    \begin{aligned}
    D F_{A(k+2),A'(k)}&=e^{BB'}F_{A(k+2)B,A'(k)B'}\\
    &-e\fud{B}{A'}\sum_{n=0}^{k-1}{{\tfrac{2}{(n+1)!}\tfrac{(k+2)!}{(k-n-1)!(k-n+1)(k+1)}}}[F_{A(n+1)B,A'(n)},F_{A(k-n+1),A'(k-n-1)}]\,.
    \end{aligned}}
\end{align}
It was assumed that the ansatz only contains linear and quadratic terms in $F$. The fact that terms cubic in $F$ vanish in \eqref{e-nabla-F1} is proved in Appendix \ref{app:sdymTruncation}. This confirms the $L_\infty$-relation in \eqref{SDYMstasheffF3} and it implies that $D F_{A(k+2),A'(k)}$ indeed truncates at quadratic order.

\paragraph{Main course, $\boldsymbol{\Psi}$-sector.} 
As was clear from the $L_\infty$-relations in \eqref{SDYMstasheff}, the non-linear extension of the $\Psi$-sector is different from the $F$-sector. The minimal ansatz for $D\Psi_{A(k),A'(k+2)}$ is slightly more involved as it reads
\begin{align}\label{ansatzpsi}
\begin{aligned}
    D\Psi_{A(k),A'(k+2)}&=e^{CC'}\Psi_{A(k)C,A'(k+2)C'}+\sum_{n=0}^{k-1}\beta_{nk}e\fud{C}{A'}[F_{A(n+1)C,A'(n)},\Psi_{A(k-n-1),A'(k-n+1)}]\\
    &+\sum_{n=0}^{k-2}\gamma_{nk}e\fud{C}{A'}[F_{A(n+2),A'(n)},\Psi_{A(k-n-2)C,A'(k-n+1)}]\,.
\end{aligned}
\end{align}
We follow the same steps as for the $F$-sector: we write the Bianchi identity for the ansatz above and as a parallel calculation we contract $e^{BB'}$ with $\Psi_{A(k+1),A'(k+3)}$ to obtain two expressions for $e^{BB'}\wedge D \Psi_{A(k)B,A'(k+2)B'}$ and compare them. This provides us with a system of recurrence relations for $\beta_{nk}$ and $\gamma_{nk}$. The details of the calculation are left for Appendix \ref{app:SDYMPsi}. The system is solved by
\begin{align*}
    \beta_{nk}&=-\tfrac{2}{(n+1)!}\tfrac{k-n+2}{k+3}\tfrac{k!}{(k-n-1)!}\,, & \gamma_{nk}&=\tfrac{2}{(n+2)!}\tfrac{n+1}{k+3}\tfrac{k!}{(k-n-2)!}\,.
\end{align*}
The full solution reads
\begin{align}\label{solpsi}\boxed{
\begin{aligned}
    D\Psi_{A(k),A'(k+2)}&=e^{CC'}\Psi_{A(k)C,A'(k+2)C'}\\
        &-e\fud{C}{A'}\sum_{n=0}^{k-1}\tfrac{2}{(n+1)!}\tfrac{k-n+2}{k+3}\tfrac{k!}{(k-n-1)!}[F_{A(n+1)C,A'(n)},\Psi_{A(k-n-1),A'(k-n+1)}]\\
        &+e\fud{C}{A'}\sum_{n=0}^{k-2}\tfrac{2}{(n+2)!}\tfrac{n+1}{k+3}\tfrac{k!}{(k-n-2)!}[F_{A(n+2),A'(n)},\Psi_{A(k-n-2)C,A'(k-n+1)}]\,.
\end{aligned}}
\end{align}
In Appendix \ref{app:sdymTruncation} we show that this solution ensures consistency of the $L_\infty$-relation in \eqref{SDYMstafhessPsi3}, i.e. the above solution does not require higher order corrections.

\paragraph{Summary.}
SDYM can be cast in the form of an $L_\infty$-algebra. This gives rise to three $L_\infty$-relations for the $F$-sector and the $\Psi$-sector of SDYM, see \eqref{SDYMstasheff}. The first of each gives rise to the free equation for $DF_{A(k+2),A'(k)}$ and $D\Psi_{A(k),A'(k+2)}$. The second $L_\infty$-relation can be solved to obtain the quadratic piece of the non-linear extension in both sectors, which are proportional to $[F,F]$ and $[F,\Psi]$, respectively. In particular, the coefficients can be found by writing down the minimal ans{\"a}tze \eqref{spin1ansatz} and \eqref{ansatzpsi} and checking their Bianchi identities. This yields two expressions for $e^{BB'}F_{A(k+2)B,A'(k)B'}$ and $e^{BB'}\Psi_{A(k)B,A'(k+2)B'}$. Comparing them gives rise to a system of recurrence relations, whose solution gives the final results \eqref{spin1sol} and \eqref{solpsi}, i.e. the boxed equations above. Furthermore, the third $L_\infty$-relation ensures that the system is closed, i.e. there are no higher order corrections. It is proved that these relation are indeed satisfied for the obtained solutions and hence the expressions we have found are the complete non-linear extensions for the two sectors.

An interesting follow up would be to consider the higher-spin extensions of SDYM \cite{Ponomarev:2017nrr,Krasnov:2021nsq} and the supersymmetric higher spin extensions constructed in \cite{Devchand:1996gv}.

\subsection{FDA, constant curvature space}
\label{sec:SDYMcurved}
As a simple modification of SDYM on Minkowski background we can consider a constant curvature background, i.e. de Sitter or anti-de Sitter spaces. The action is the same. Let us first recall that the free Maxwell equations on a constant curvature background rewritten as an FDA read \cite{Vasiliev:1986td}
\besubeqs \label{Maxw}
\begin{align}\label{eqMaxwAA}
    dA&=H^{BB}F_{BB} +\epsilon H^{B'B'} \Psi_{B'B'} \,, \\
\label{eqMaxwBA}
    \nabla F^{A(k+2),A'(k)}&=e_{BB'}F^{A(k+2)B,A'(k)B'} +k(k+2)\Lambda e^{AA'} F^{A(k+1), A'(k-1)} \,, \\
\label{eqMaxwCA}
    \nabla \Psi^{A(k),A'(k+2)}&=e_{CC'}\Psi^{A(k)C,A'(k+2)C'}+k(k+2)\Lambda e^{AA'} \Psi^{A(k-1),A'(k+1)} \,.
\end{align}
\esubeqs
The only difference is the presence of new $e^{AA'}$-terms that are consistent on their own and do not require any other modifications. It is also convenient to set $\Lambda=1$ in what follows. The $L_\infty$-algebra for SDYM on a constant background is given by
\begin{align*}
    d e^{AA'}&= \omega\fud{A}{B}\wedge e^{BA'}+\omega\fud{A'}{B'}\wedge e^{A B'} \,, \\
    d\omega^{AB}&= \omega\fud{A}{C}\wedge \omega^{BC}+  H^{AB} \,, \\
    d\omega^{A'B'}&= \omega\fud{A'}{C'}\wedge \omega^{B'C'}+  H^{A'B'} \,, \\
    dA&=  AA + H_{BB}F^{BB} \,, \\
    dF&= l_2(\omega,F) + l_2(A,F)+l_2(e,F)+\tilde{l}_2(e,F)+l_3(e,F,F) \,, \\
    d\Psi&= l_2(\omega,\Psi) + l_2(A,\Psi)+l_2(e,\Psi)+\tilde{l}_2(e,\Psi) +l_3(e,F,\Psi)\,,
\end{align*}
where $\tilde{l}_2$ encodes the gravitational correction to the free equations \eqref{eqMaxwBA} and \eqref{eqMaxwCA}. The contributions $$l_2(\omega,F)= (k+2)\omega\fud{A}{B}F^{A(k+1)B,A'(k)}+k\omega\fud{A'}{B'}F^{A(k+2),B'A'(k-1)}$$ and $l_2(A,F)=[A,F]$ (and similarly for $\Psi$) can be absorbed into the covariant derivative $D=\nabla-[A,\bullet]$. As a result, the relations can be rewritten as
\begin{align*}
    \nabla e^{AA'}&= 0 \,, \\ 
    dA&= AA + H_{BB}F^{BB} \,, \\
    DF&= l_2(e,F)+\tilde{l}_2(e,F)+l_3(e,F,F) \,, \\
    D\Psi&= l_2(e,\Psi)+\tilde{l}_2(e,\Psi)+l_3(e,F,\Psi)\,.
\end{align*}
As different from $\nabla^2=0$ in flat space, in a constant curvature background we have for any spin-tensor $T^{A(n),A'(m)}$
\begin{align*}
    \nabla^2T^{A(n),A'(m)}&=-n  H\fud{A}{B}T^{A(n-1)B,A'(m)}-m  H\fud{A'}{B'}T^{A(n),A'(m-1)B'} \,.
\end{align*}
The $L_\infty$-relations of the sought for $L_\infty$-algebra read
\besubeqs
    \begin{align}
    -[H_{BB}F^{BB},F]+l_2(e,DF)+\tilde{l}_2(e,DF)+l_3(e,DF,F)+l_3(e,F,DF)&\equiv0 \,, \\
    -[H_{BB}F^{BB},\Psi]+l_2(e,D\Psi)+\tilde{l}_2(e,D\Psi)+l_3(e,DF,\Psi)+l_3(e,F,D\Psi)&\equiv0 \,.
    \end{align}
\esubeqs
Since $\tilde{l}$ can be viewed as a deformation of the previously found FDA, all terms without $\tilde{l}$ vanish already. The remaining nontrivial relations read
\besubeqs\label{SDYMstasheffLambda}
\begin{align} 
    \tilde{l}_2(e,l_3(e,F,F))+l_3(e,\tilde{l}_2(e,F),F)+l_3(e,F,\tilde{l}_2(e,F))&=0 \,, \label{stasheffLambdaF} \\
    \tilde{l}_2(e,l_3(e,F,\Psi))+l_3(e,\tilde{l}_2(e,F),\Psi)+l_3(e,F,\tilde{l}_2(e,\Psi))&=0 \label{stasheffLambdaPsi} \,,
\end{align}
\esubeqs
where we ignore terms quadratic in the cosmological constant. These relations are satisfied automatically. 
A proof of this given in Appendix \ref{app:sdymLambda}. Consequently, on a constant curvature gravitational background we obtain
\besubeqs
    \begin{align} \label{curvedFsol}\boxed{
        \begin{aligned}
            D& F_{A(k+2),A'(k)}=e^{BB'}F_{A(k+2)B,A'(k)B'}+k(k+2)e_{AA'}F_{A(k+1),A'(k-1)}\\
            &\qquad-e\fud{B}{A'}\sum_{n=0}^{k-1}\tfrac{2}{(n+1)!}\tfrac{(k+2)!}{(k-n-1)!(k-n+1)(k+1)}[F_{A(n+1)B,A'(n)},F_{A(k-n+1),A'(k-n-1)}]\,,
        \end{aligned}}
    \end{align}
    \begin{align} \label{curvedPsisol}\boxed{
        \begin{aligned}
        D\Psi_{A(k),A'(k+2)}&=e^{CC'}\Psi_{A(k)C,A'(k+2)C'}+k(k+2)e_{AA'}\Psi_{A(k-1),A'(k+1)}\\
        &-e\fud{C}{A'}\sum_{n=0}^{k-1}\tfrac{2}{(n+1)!}\tfrac{k-n+2}{k+3}\tfrac{k!}{(k-n-1)!}[F_{A(n+1)C,A'(n)},\Psi_{A(k-n-1),A'(k-n+1)}]\\
        &+e\fud{C}{A'}\sum_{n=0}^{k-2}\tfrac{2}{(n+2)!}\tfrac{n+1}{k+3}\tfrac{k!}{(k-n-2)!}[F_{A(n+2),A'(n)},\Psi_{A(k-n-2)C,A'(k-n+1)}]\,.
        \end{aligned}}
    \end{align}
\esubeqs

\paragraph{Summary.} We constructed the $L_\infty$-algebra of SDYM on a constant curvature background and derived the corresponding $L_\infty$-relations. The free Maxwell equations on a constant curvature background in terms of an FDA, \eqref{Maxw}, are well-known in the literature and solve the first $L_\infty$-relation of both the $F$-sector and $\Psi$-sector. In section \ref{sec:flatSDYM} we computed the non-linear extension of $DF_{A(k+2),A'(k)}$ and $D\Psi_{A(k),A'(k+2)}$ on a flat background. In the second $L_\infty$-relation of each sector we see an interplay between the gravitational contribution of the free equations and the non-linear extension on flat space. We demonstrated that the second $L_\infty$-relation for both sectors decomposes into the flat space $L_\infty$-relation and a new relation containing the gravitational contributions in such a way that the latter does not contribute to the quadratic order in $DF_{A(k+2),A'(k)}$ and $D\Psi_{A(k),A'(k+2)}$. The third $L_\infty$-relation then contains no gravitational contribution and remains satisfied. The complete non-linear extension of both sectors are only modified in the linear terms according to the free equations and are shown in the boxed equation \eqref{curvedFsol} and \eqref{curvedPsisol} above.

\section{SDGR}
\label{sec:SDGR}

\subsection{Action, initial data}
\label{sec:}
Self-dual gravity with vanishing cosmological constant can be formulated with the help of two fields \cite{Krasnov:2021cva}: one-form $\omega^{A'B'}$ and zero-form $\Psi^{A'B'C'D'}$. The action reads
\begin{align}\label{flsd}
    \int \Psi^{A'B'C'D'}\wedge d\omega_{A'B'} \wedge d\omega_{C'D'} \,.
\end{align}
The equations of motion are ($F^{A'B'}=d\omega^{A'B'}$)
\begin{align}\label{fleq}
    F_{(A'B'} \wedge F_{C'D')}&=0 \,, &  d\Psi^{A'B'C'D'}\wedge F_{A'B'}&=0 \,.
\end{align}
One-form $\omega^{A'B'}$ looks like the anti-self-dual part of the Lorentz spin-connection, but it is not. The curvature $F_{A'B'}$ for $\omega^{A'B'}$ lacks the "$\omega\omega$"-part. Nevertheless, this interpretation is not very far from the reality since action \eqref{flsd} can be understood as a limit of that for self-dual gravity with cosmological constant \cite{Krasnov:2016emc}. In the latter $F^{A'B'}=d\omega^{A'B'}-\omega\fud{A'}{C'}\wedge \omega^{C'B'}$ is the canonical one and the limit is to drop the $\omega\omega$-part. 

Minkowski space is a special solution of \eqref{fleq}: $\omega_0^{A'A'}=x\fdu{C}{A'} dx^{CA'}$ such that $d\omega_0^{A'B'}=H^{A'B'}$, where $H^{A'B'}$ is built from the Minkowski's space vierbein $e^{AA'}=dx^{AA'}$, $H^{A'B'}\equiv e\fdu{C}{A'}\wedge e^{CB'}$ and its conjugate is $H^{AB}\equiv e\fud{A}{C'}\wedge e^{BC'}$. One can easily write down the first few equations of the FDA that corresponds to variational equations \eqref{fleq}:\footnote{As a side remark, let us write the curvature for $so(3,2)\sim sp(4)$, which is relevant for anti-de Sitter space (they correspond to Lorentz generators $L_{A'A'}$, $L_{AA}$ and to translations $P_{AA'}$):
\besubeqs
\begin{align*}
    d \omega^{AA}-\omega\fud{A}{C}\wedge\omega^{CB}- e\fud{A}{B'}\wedge e^{AB'}&=R^{AA}\,,\\
    d e^{AA'}-\omega\fud{A'}{B'}\wedge e^{AB'}-\omega\fud{A}{B}\wedge e^{BA'} &=T^{AA'}\,,\\
    d \omega^{A'A'}-\omega\fud{A'}{C'}\wedge\omega^{C'B'}- e\fdu{B}{A'}\wedge e^{BA'}&=R^{A'A'}\,,
\end{align*}
\esubeqs
The gauge algebra for the SDGR with zero scalar curvature can be understood as a limit of $so(3,2)$-algebra where $L_{A'A'}$ become abelian \cite{Krasnov:2021cva}. 
}
\besubeqs\label{gaugesect}
\begin{align} \label{RandPsi}
    &\qquad\begin{aligned}
     d \omega^{A'A'}&=e\fdu{B}{A'}\wedge e^{BA'}\,,\\
    d e^{AA'}&=\omega\fud{A}{B}\wedge e^{BA'} \,,\\
    d \omega^{AA}&=\omega\fud{A}{C}\wedge\omega^{CA}+H_{MM}C\fud{MMA}{B} \,, 
    \end{aligned}\\
    &\,\, d\Psi^{A'A'A'A'}= e_{BB'}\Psi^{B,A'A'A'A'B'}\,. \label{firstpsi}
\end{align}
\esubeqs
The main idea is to identify the right gauge algebra \cite{Krasnov:2021cva}. This is the starting for constructing the $L_\infty$-algebra. The first equation of \eqref{RandPsi} implies that the gravitational degrees of freedom fully reside in the anti-self-dual part. The last equation of \eqref{RandPsi} identifies the only nonvanishing part of the curvature with the self-dual Weyl tensor $C^{ABCD}$, $R\fud{A}{B}=H_{MM}C\fud{MMA}{B} $. As a result one obtains a Bianchi identity for $R\fud{A}{B}$. Eq. \eqref{firstpsi} introduces a new field $\Psi^{A,A'B'C'D'E'}$, which parameterizes the first derivative of $\Psi$ and is contained in the on-shell jet of $\Psi^{A'B'C'D'}$. Similarly to SDYM we aim to find a completion of \eqref{RandPsi} and we need to define an infinite set of coordinates on $\mathcal{N}$ and $Q$ such that $QQ=0$. 

\paragraph{Coordinates, on-shell jet.} Coordinates on supermanifold $\mathcal{N}$ coincide with those of the free massless spin-two field, i.e. with \cite{Vasiliev:1986td} and \cite{penroserindler}. Indeed, the set of one-forms turned out to be the same, while the zero-forms begin with (anti)-self-dual components of Weyl tensor and are just the on-shell nontrivial derivatives of those. Therefore, the coordinates on $\mathcal{N}$ are: degree-one $\omega^{AB}$, $e^{AA'}$ and $\omega^{A'B'}$; degree-zero $C^{A(k+4),A'(k)}$ and $\Psi^{A(k+4),A'(k)}$, $k=0,1,2,...$. A similar discussion follows as for SDYM. In particular, the free equations for helicity $\pm 2$ fields are \cite{Penrose:1965am}
\begin{align*}
    \nabla\fud{A}{B'}\Psi^{A'B'C'D'}&=0 \,, & \nabla\fdu{A}{B'}C^{ABCD}&=0\,,
\end{align*}
and can be rewritten in the FDA form as \cite{Vasiliev:1986td}
\begin{align}\label{freespintwo}
    \nabla C^{A(k+4),A'(k)}&=e_{CC'}C^{A(k+4)C,A'(k)C'} \,, & \nabla \Psi^{A(k),A'(k+4)}&=e_{CC'}\Psi^{A(k)C,A'(k+4)C'} \,.
\end{align}
One needs to supplement these equations with the free limit of \eqref{gaugesect}. Our problem is to find a nonlinear completion of \eqref{freespintwo} that is consistent with \eqref{gaugesect}.

\paragraph{General form.} 
The supermanifold $\mathcal{N}$ has coordinates
\begin{align*}
    \mathcal{N}&: && 
    \begin{aligned}
        1&: \omega^{A'B'}\,,e^{AA'}\,, \omega^{AB} \,,\\\
        0&: C^{A(k+4),A'(k)}\,, \Psi^{A(k),A'(k+4)}\,, k=0,1,2,...
    \end{aligned}
\end{align*}
Now, we try to reformulate the theory in the $L_\infty$-form. Given the data above and our desire to truncate the FDA at $l_3(\bullet,\bullet,\bullet)$, we write 
\begin{align*}
     d \omega^{A'A'}&=e\fdu{B}{A'}\wedge e^{BA'}\,,\\
    d e^{AA'}&=\omega\fud{A}{B}\wedge e^{BA'} \,,\\
    d \omega^{AA}&=\omega\fud{A}{C}\wedge\omega^{CA}+H_{MM}C\fud{MMA}{B} \,, \\
    dC&= l_2(\omega,C)+l_2(e,C)+l_3(e,C,C) \,, \\
    d\Psi&= l_2(\omega,\Psi)+l_2(e,\Psi) +l_3(e,C,\Psi)\,.
\end{align*}
We define the covariant derivative $\nabla=d-\omega$, which lacks the $\omega^{A'B'}$-part. For an arbitrary spin-tensor $T^{A(n),A'(m)}$ we get
\begin{align} \label{nablasquared}
    \nabla^2T^{A(n),A'(m)}=-nH_{MM}C\fud{MMA}{B}T^{BA(n-1),A'(m)} \,.
\end{align}
The covariant derivative allows one to absorb the terms $l_2(\omega,C)$ and $l_2(\omega,\Psi)$ and we can write
    \begin{align*}
        \nabla C&=l_2(e,C)+l_3(e,C,C) \,, &
        \nabla \Psi&=l_2(e,\Psi)+l_3(e,C,\Psi)\,.
    \end{align*}
This gives rise to the $L_\infty$-relations for SDGR, which read
\begin{align*}
    \begin{aligned}
        -(k+4) H_{MM}C\fud{MMA}{B}C^{A(k+3)B,A'(k)}+l_2(e,\nabla C)+l_3(e,\nabla C,C)+l_3(e,C,\nabla C)&=0 \,, \\
        -k H_{MM}C\fud{MMA}{B}\Psi^{A(k-1)B,A'(k+4)}+l_2(e,\nabla\Psi)+l_3(e,\nabla C,\Psi)+l_3(e,C,\nabla\Psi)&=0\,,
    \end{aligned}
\end{align*}
and decompose into
\besubeqs \label{SDGRstasheff}
    \begin{align} 
        &l_2(e,l_2(e,C))=0 \label{stasheffC1} \,, \\
           &l_3(e,l_3(e,C,C),C)+l_3(e,C,l_3(e,C,C))=0 \label{stasheffC3} \,, \\
       & l_2(e,l_2(e,\Psi))=0 \label{stasheffSDGRPsi1} \,, \\     
       & l_3(e,l_3(e,C,C),\Psi)+l_3(e,C,l_3(e,C,\Psi))=0 \label{stasheffSDGRPsi3} \,,\\
        &\begin{aligned}
            -n H_{MM}C\fud{MMA}{B}C^{A(k+3)B,A'(k)}&+l_2(e,l_3(e,C,C))       \\
            &+l_3(e,l_2(e,C),C)+l_3(e,C,l_2(e,C))=0  \,, \label{stasheffC2}
        \end{aligned}\\
        &\begin{aligned}
           -nH_{MM}C\fud{MMA}{B}\Psi^{A(k-1)B,A'(k+4)}&+l_2(e,l_3(e,C,\Psi))\\&+l_3(e,l_2(e,C),\Psi)+l_3(e,C,l_2(e,\Psi))=0\,. \label{stasheffSDGRPsi2} 
        \end{aligned}
    \end{align}
\esubeqs

\subsection{FDA}
\label{sec:SDGRflat}
\paragraph{Appetizer.}
Let us first illustrate our approach by presenting the source of the non-linear extension with an explicit example. We follow roughly the same steps as for SDYM, though some subtle differences arise. The most important ones come from the commutativity of the $C$'s and the additional contraction of unprimed indices that we will see shortly.

The Bianchi identity for the curvature, $\nabla R_{AA}=0$ implies
\begin{align*}
    \nabla C_{AAAA}=e^{BB'}C_{AAAB,B'} \,.
\end{align*}
Its own Bianchi identity via \eqref{nablasquared} imposes
\begin{align*}
    \nabla^2C^{AAAA}=-e^{BB'}\wedge\nabla C_{AAAAB,B'}=4H^{BB}C\fdu{ABB}{D}C_{AAAD} \,.
\end{align*}
We need to construct an ansatz for $\nabla C_{AAAAA,A'}$. Commutativity of the $C$'s and the Fierz identity allow us to construct the minimal ansatz as
\begin{align} \label{eq:nablaC}
    \nabla C_{AAAAA,A'}=e^{CC'}C_{AAAAAB,A'C'}+a_{01}e\fud{C}{A'}C\fdu{AAC}{D}C_{AAAD}\,.
\end{align}
Contracting the ansatz with $e^{BB'}$ yields
\begin{align*}
    \begin{aligned}
        e^{BB'}\wedge\nabla C_{AAAAB,B'}&=-\tfrac{2a_{01}}{5}H^{BB}C\fdu{ABB}{D}C_{AAAD}-\tfrac{3a_{01}}{5}H^{BB}C\fdu{AAB}{D}C_{AABD}\\
        &=-\tfrac{2a_{01}}{5}H^{BB}C\fdu{ABB}{D}C_{AAAD}\,.
    \end{aligned}
\end{align*}
One term is dropped, as commuting the two $C$'s and raising/lowering the contracted indices tells us that this term vanishes. Comparing the result with \eqref{eq:nablaC} yields the solution
\begin{align*}
    \nabla C_{AAAAA,A'}=e^{CC'}C_{AAAAAC,A'C'}+10e\fud{C}{A'}C\fdu{AAC}{D}C_{AAAD}\,.
\end{align*}
The procedure that we have followed is a practical realisation of solving the $L_\infty$-relation \eqref{stasheffC2}. This procedure will be generalized next.

\paragraph{Main course, $\boldsymbol{C}$-sector.} Using the same criteria as before we propose the minimal ansatz
\begin{align*}
    \nabla C_{A(k+4),A'(k)}=e^{CC'}C_{A(k+4)C,A'(k)C'}+\sum_{n=0}^{k-1}a_{nk}e\fud{C}{A'}C\fdud{A(n+2)C}{D}{,A'(n)}C_{A(k-n+2)D,A'(k-n-1)}\,.
\end{align*}
Taking another derivative leads to
\begin{align} \label{nablasquaredC}
    \begin{aligned}
        \nabla^2C_{A(k+4),A'(k)}&=(k+4)H^{BB'}C\fdu{ABB}{D}C_{A(k+3)D,A'(k)}=-e^{CC'}\wedge\nabla C_{A(k+4)C,A'(k)C'}\\
        &-\sum_{n=1}^{k-1}a_{nk}e\fud{C}{A'}\wedge\nabla C\fdud{A(n+2)C}{D}{,A'(n)}C_{A(k-n+2)D,A'(k-n-1)}\\
        &-\sum_{n=0}^{k-2}a_{nk}e\fud{C}{A'}\wedge C\fdud{A(n+2)C}{D}{,A'(n)}\nabla C_{A(k-n+2)D,A'(k-n-1)}\,.
    \end{aligned}
\end{align}
Considering only terms quadratic in $C$ yields
\begin{align} \label{eC}
    \begin{aligned}
        e^{CC'}\wedge\nabla C_{A(k+4)C,A'(k)C'}&=-(k+4)H^{BB}C\fdu{ABB}{D}C_{A(k+3)D,A'(k)}\\
        &-\tfrac{1}{2}H^{BB}\sum_{n=0}^{k-1}a_{nk}C\fdud{A(n+2)BB}{D}{,A'(n+1)}C_{A(k-n+2)D,A'(k-n-1)}\\
        &-\tfrac{1}{2}H^{BB}\sum_{n=0}^{k}(\tfrac{a_{nk}}{2}-\tfrac{a_{(k-n)k}}{2})C\fdud{A(n+2)B}{D}{,A'(n)}C_{A(k-n+2)BD,A'(k-n)}\\
        &-\tfrac{1}{2}H\fdu{A'}{B'}\sum_{n=0}^{k-1}a_{nk}C\fdud{A(n+2)B}{D}{,A'(n)}C\fdud{A(k-n+2)}{B}{D,A'(k-n-1)B'}\,,
    \end{aligned}
\end{align}
where in the third line we made the anti-commuting property of the $C$'s explicit, together with the anti-symmetry of the spinorial inner product. At the same time we contract $e^{BB'}$ with $\nabla C_{A(k+5),A'(k+1)}$ to obtain
\begin{align*}
    \begin{aligned}
    e^{BB'}&\wedge\nabla C_{A(k+4)B,A'(k)B'}=-H^{BB}\tfrac{k+2}{(k+5)(k+1)}a_{0(k+1)}C\fdu{ABB}{D}C_{A(k+3)D,A'(k)}\\
    &-\tfrac{1}{2}H^{BB}\sum_{n=0}^{k-1}\tfrac{(k+2)(n+3)}{(k+5)(k+1)}a_{(n+1)(k+1)}C\fdud{A(n+2)BB}{D}{,A'(n+1)}C_{A(k-n+2)D,A'(k-n-1)}\\
    &-\tfrac{1}{4}H^{BB}\sum_{n}^{k}(\tfrac{(k+2)(k-n+3)}{(k+5)(k+1)}a_{n(k+1)}-\tfrac{(k+2)(n+3)}{(k+5)(k+1)}a_{(k-n)(k+1)})\\
    &\times C\fdud{A(n+2)B}{D}{,A'(n)}C_{A(k-n+2)BD,A'(k-n)}\\
    &-\tfrac{1}{2}H\fdu{A'}{B'}\sum_{n=0}^{k-1}(\tfrac{(k-n)(n+3)}{(k+5)(k+1)}a_{(k-n)(k+1)}+\tfrac{(k-n)(k-n+3)}{(k+5)(k+1)}a_{n(k+1)})\\
    &\times C\fdud{A(n+2)B}{D}{,A'(n)}C\fdud{A(k-n+2)}{B}{D,A'(k-n-1)B'} \,.
    \end{aligned}
\end{align*}
Comparing this expression with \eqref{eC} brings about the following system of recurrence relations:
    \begin{align*}
        0&=a_{0k}-\tfrac{(k+4)(k+3)k}{k+1} \,, \\
        0&=a_{(n+1)(k+1)}-\tfrac{(k+5)(k+1)}{(k+2)(n+3)}a_{nk} \,, \\
        0&=\tfrac{(k+2)(k-n+3)}{(k+5)(k+1)}a_{n(k+1)}-\tfrac{(k+2)(n+3)}{(k+5)(k+1)}a_{(k-n)(k+1)}-a_{nk}+a_{(k-n)n} \,, \\
        0&=a_{nk}-\tfrac{(k-n)(n+3)}{(k+5)(k+1)}a_{(k-n)(k+1)}-\tfrac{(k-n)(k-n+3)}{(k+5)(k+1)}a_{n(k+1)} \,.
    \end{align*}
This over-determined system is solved by
\begin{align*}
    a_{nk}=\tfrac{2}{(n+2)!}\tfrac{(k+4)!(k-n)}{(k-n+2)!(k+1)}
\end{align*}
and the full solution reads\footnote{A closely related problem was addressed in \cite{Vasiliev:1989xz}, which is to find an FDA form of the full gravity to the next to the leading order (the problem to find the complete minimal model for gravity does not seem to admit a solution in a closed form, even though it does always exist as a matter of principle). It would be interesting to understand what \cite{Vasiliev:1989xz} describes since it does not coincide with the FDA of SDGR with (non)-vanishing cosmological constant. The physical degrees of freedom are the same though.}
\begin{align} \label{Csol} \boxed{
    \begin{aligned}
        \nabla C_{A(k+4),A'(k)}&=e^{CC'}C_{A(k+4)C,A'(k)C'}\\
        &+\sum_{n=0}^{k-1}\tfrac{2}{(n+2)!}\tfrac{(k+4)!(k-n)}{(k-n+2)!(k+1)}e\fud{C}{A'}C\fdud{A(n+2)C}{D}{,A'(n)}C_{A(k-n+2)D,A'(k-n-1)}\,.
    \end{aligned}}
\end{align}
In appendix \ref{app:sdgrTruncation} we prove that this solution is complete, i.e. no higher order terms arise.

\paragraph{Main course, $\boldsymbol{\Psi}$-sector.} For the $\Psi$-sector we follow a similar approach. The minimal ansatz reads
\begin{align} \label{eq:SDGRPsiansatz}
    \begin{aligned}
        \nabla\Psi_{A(k),A'(k+4)}&=e^{CC'}\Psi_{A(k)C,A'(k+1)C'}+\sum_{n=0}^{k}b_{nk}e\fud{C}{A'}C\fdud{A(n+2)C}{D}{,A'(n)}\Psi_{A(k-n-2)D,A'(k-n+3)}\\
        &+\sum_{n=0}^{k}c_{nk}e\fud{C}{A'}C\fdud{A(n+3)}{D}{,A'(n)}\Psi_{A(k-n-3)CD,A'(k-n+3)}\,.
    \end{aligned}
\end{align}
The details of the calculations are left to Appendix \ref{app:SDGRPsi}, but the approach is as follows: we take the covariant derivative of the ansatz above. We also contract $e^{BB'}$ with $\nabla\Psi_{A(k+1),A'(k+5)}$. Both will give us an expression for $e^{BB'}\wedge\nabla\Psi_{A(k)B,A'(k+4)B'}$ and we compare them. This results in a system of recurrence relations, which is solved by
    \begin{align*}
        b_{nk}&=\tfrac{2}{(n+2)!}\tfrac{k!}{(k-n-2)!}\tfrac{k-n+4}{k+5} \,, &
        c_{nk}&=-\tfrac{2}{(n+2)!}\tfrac{k!}{(k-n-3)!}\tfrac{n+1}{(k+5)(n+3)}\,,
    \end{align*}
and the solution in the $\Psi$-sector reads
\begin{align} \label{psisolSDGR} \boxed{
    \begin{aligned}
        \nabla\Psi_{A(k),A'(k+4)}&=e^{CC'}\Psi_{A(k)C,A'(k+1)C'}\\
        &+\sum_{n=0}^{k}\tfrac{2}{(n+2)!}\tfrac{k!}{(k-n-2)!}\tfrac{k-n+4}{k+5}e\fud{C}{A'}C\fdud{A(n+2)C}{D}{,A'(n)}\Psi_{A(k-n-2)D,A'(k-n+3)}\\
        &-\sum_{n=0}^{k}\tfrac{2}{(n+2)!}\tfrac{k!}{(k-n-3)!}\tfrac{n+1}{(k+5)(n+3)}e\fud{C}{A'}C\fdud{A(n+3)}{D}{,A'(n)}\Psi_{A(k-n-3)CD,A'(k-n+3)}\,.
    \end{aligned}}
\end{align}
We prove in Appendix \ref{app:sdgrTruncation} that $\nabla\Psi_{A(k),A'(k+4)}$ is consistent as it is and does not require higher order terms. 

\paragraph{Summary.}
We rewrote SDGR as an $L_\infty$-algebra. This gives rise to three $L_\infty$-relations for the $C$-sector and the $\Psi$-sector, see \eqref{SDGRstasheff}. Solving the first relation of each sector yields the free equations for $\nabla C^{A(k+4),A'(k)}$ and $\nabla \Psi^{A(k),A'(k+4)}$. We constructed a minimal ansatz for a non-linear extension of the free equations and used the second $L_\infty$-relation to determine its structure. The results are shown in the boxes expressions above, \eqref{Csol} and \eqref{psisolSDGR}. The third $L_\infty$-relation is found to be satisfied for the obtained solutions, which implies that the minimal ansatz is sufficient to solve the whole system.

An interesting followup of this project is construct FDA for SDGR in the constant-curvature background. The action of this theory \cite{Krasnov:2016emc} is even more natural 
\begin{align*}
    \int \Psi^{A'B'C'D'}\wedge F_{A'B'} \wedge F_{C'D'}\,,
\end{align*}
where $F^{A'B'}=d\omega^{A'B'}-\omega\fud{A'}{C'}\wedge \omega^{C'B'}$. However, it is more nonlinear, featuring quartic terms (the quintic one vanishes). A simpler problem is to consider the higher spin extensions of SDGR \cite{Ponomarev:2017nrr,Krasnov:2021nsq} with vanishing cosmological constant.

\section{Conclusions and Discussion}
The present paper is the first in a series of papers where we plan to construct minimal models of various field theories, including some examples of higher spin gravities. Since every (gauge) field theory defines and is defined by its minimal model, a certain $L_\infty$-algebra, our general motivation is to first understand how various properties of field theories, e.g. integrability, asymptotic symmetries, conserved charges, actions, anomalies etc., can be understood in the known cases and derived from this $L_\infty$-algebra in the cases where this information is yet unavailable. For example, it would be interesting to understand the Ward construction of Yang-Mills instantons \cite{Ward:1977ta} from the $L_\infty$ point of view.

As we have reviewed in section \ref{sec:FDA}, the minimal model can naturally be associated to any gauge theory and it is the smallest $L_\infty$-algebra that captures all local BRST cohomology of this field theory. However, the minimal model is usually difficult to construct explicitly. Apart from this paper, the only available examples where minimal models were explicitly constructed are (a) Chern-Simons theory, which is just $dA=AA$ and, for that reason, is hard to consider this as a genuine example of a minimal model (nevertheless, this toy model was quite useful to prove that all matter-free higher spin gravities in $3d$ are of Chern-Simons form \cite{Grigoriev:2020lzu}); (b) another example is discussed in \cite{Vasiliev:1989xz} and is closely related to the SDGR FDA of the present paper. It is tempting to argue that minimal models can explicitly be constructed only for theories that feature some kind of (hidden) simplicity, e.g. they are integrable like Chern-Simons theory, SDYM and SDGR.

Some obvious future directions include: (a) self-dual gravity with cosmological constant \cite{Krasnov:2016emc}; (b) higher spin extensions of SDYM and SDGR \cite{Ponomarev:2017nrr,Krasnov:2021nsq}; (c) the supersymmetric higher spin extensions of \cite{Devchand:1996gv}; Chiral higher spin gravity \cite{Metsaev:1991mt,Metsaev:1991nb,Ponomarev:2016lrm,Ponomarev:2017nrr,Skvortsov:2018jea,Skvortsov:2020wtf,Skvortsov:2020gpn}. 

All physically relevant local information about a given theory is encoded in its minimal model via the $Q$-cohomology. For example, conserved  charges, actions correspond to $H(Q)$ with values in the trivial module. A more complicated example is the presymplectic structure $\Omega_{\aA \aB}\, \delta\Phi^\aA \wedge \delta \Phi^\aB$ which is a two-form on the field space and is a degree $(d-1)$ form from the space-time point of view. It corresponds to $H(Q,\Lambda^2( \mathcal{N}))$, where the action of $Q$ is understood as Lie derivative $L_Q$ along $Q$ that is defined canonically on $(p,q)$-tensors on $\mathcal{N}$, see \cite{Alkalaev:2013hta,Grigoriev:2016wmk,Sharapov:2020quq,Sharapov:2021drr} for more detail and examples. As the last example, $Q$-cohomology with values in vector fields, $H(Q,T^{1,0}(\mathcal{N}))$, is responsible for deformations of $Q$ itself, i.e. it classifies possible interactions. It is worth noting that $Q$-cohomology can often be computed without having to know the minimal model explicitly. The latter is an additional bonus that should be a signal of integrability.

\section*{Acknowledgments}
\label{sec:Aknowledgements}
We are grateful to Maxim Grigoriev, Yannick Herfray, Kirill Krasnov and Alexey Sharapov for useful discussions. This project has received funding from the European Research Council (ERC) under the European Union’s Horizon 2020 research and innovation programme (grant agreement No 101002551). The work was partially supported by the Fonds de la Recherche Scientifique --- FNRS under Grant No. F.4544.21.

\appendix

\section{Notation}\label{app:notation}
The most important conventions and definitions that were used in the main text are introduced here. A short discussion on the spinor formalism is given, a more elaborate treatment can be found in \cite{penroserindler}.

It is useful for $4d$ theories to express all space-time indices in terms of spinor indices, using $so(3,1)\sim sl(2,\mathbb{C})$. This isomorphism allows to map $4$-dimensional space-time vectors to $2\times2$ Hermitian matrices, which can be extended to tensors. As basis for the $2\times2$ Hermitian matrices in flat space-time we choose the Pauli matrices and the unit matrix $\sigma^\mu_{AB'}=(1\!\!1,\sigma^i)$. The Greek letters run over space-time indices, whereas the lower case Latin letters are space indices and capitals are the matrix indices. The Pauli matrices satisfy $\mathrm{Tr}\,\sigma^i=0$ and $\{\sigma^\mu,\sigma^\nu\}_{AA'}=2\eta^{\mu\nu}1\!\!1_{AA'}$, with $\eta^{\mu\nu}=\text{diag}(-1,1,1,1)$ the Minkowski metric. We define
\begin{align*}
    x_{AA'}=x^\mu(\sigma_{\mu})_{AA'}=\begin{pmatrix}
    x^0+x^3 & x^1-ix^2\\
    x^1+ix^2 & x^1-x^3
    \end{pmatrix} \,,
\end{align*}
which is Hermitian. We also introduce a dual set $(\bar{\sigma}^{\mu})^{   AA'}=(\mathbf{1},-\sigma^i)$, such that
\begin{align*} 
    v^\mu=-\tfrac{1}{2}x_{AA'}\bar{\sigma}^{\mu A'A} \,.
\end{align*}
The two sets are related by $\sigma_\mu^{AA'}=\bar{\sigma}_\mu^{A'A}$. We also introduce raising and lowering rules for the primed (and similarly for unprimed indices):
\begin{align*}
    y_A&=y^B\epsilon_{BA},& y^A&=\epsilon^{AB}y_B \,.
\end{align*}
The inner product in spinor indices is defined as $(xy)=x^Ay^B\epsilon_{AB}=x_Ay^A=-x^Ay_A$. We define the $\epsilon$'s as
\begin{align*}
    \epsilon^{AB}=\epsilon^{A'B'}=i(\sigma^2)^{AB}=\begin{pmatrix}
    0 & 1\\
    -1 & 0
    \end{pmatrix}
\end{align*}
and their inverse $-\epsilon_{AB}=(-\sigma^2)^{-1}=-i\sigma^2$. $\epsilon_{AB}$ is anti-symmetric and $\epsilon_{AC}\epsilon^{BC}=\delta_{A}^{\hspace{5pt}B}$. Inner products in spinor indices look slightly different from inner products in space-time indices:
\begin{align*}
    x_{AA'}y^{AA'}&=-2x_\mu y^\mu\,, & z_{A}z^{A}&=z^{A}z^{B}\epsilon_{AB}=0\,.
\end{align*}
Any bi-spinor $T_{AB}$ can be decomposed into symmetric and anti-symmetric parts:
\begin{align} \label{eq:spinordecomp}
T_{AB}=\tfrac{1}{2}(T_{AB}+T_{BA}+T_{AB}-T_{BA})=T_{(AB)}+\tfrac{1}{2}\epsilon_{AB}T_C^{\hspace{5pt}C} \,,
\end{align}
where $T_{(AB)}=\tfrac{1}{2!}(T_{AB}+T_{BA})$ denotes the symmetric part of $T_{AB}$. From now on, we will use the convention that if a tensor carries identical indices, it is implied that the tensor is symmetric in them, e.g. $T^{AA}=\tfrac{1}{2!}(T^{A_1A_2}+T^{A_2A_1})$ and in a more condensed notation, $T^{A(n)}$ is symmetrized over the $n$ indices in a similar fashion. Tensors can carry two types of unrelated indices, primed and unprimed. In the most general case we write $T^{A(m),A'(n)}=\tfrac{1}{m!n!}\sum_{permutations}T^{A_1\dots A_m,A'_1\dots A'_n}$.
\\
An object that we will often use is the vierbein $e^{AA'}\equiv e_\mu^{AA'}dx^\mu$, which is a one-form. A direct consequence of the decomposition of \eqref{eq:spinordecomp} leads to the important identity
\begin{align} \label{e-wedge-e}
    e_{AA'}\wedge e_{BB'}=\tfrac{1}{2}(\epsilon_{A'B'}H_{AB}+\epsilon_ {AB}H_{A'B'}) \,,
\end{align}
where $H_{AB}=e_{AC'}\wedge e\fdu B{C'}$ and $H_{A'B'}=e_{CA'}\wedge e\fud C{B'}$. This identity allows one, for example, to rewrite the Yang-Mills field strength in terms of its (anti)-self-dual parts
\begin{align*}
    F=F_{AA'|BB'}e^{AA'}\wedge e^{BB'}=H^{BB}F_{BB}+H^{B'B'}F_{B'B'}\,,
\end{align*}
with $F_{AB}=\tfrac{1}{2}F\fdu{AC'|B}{C'}$ and $F_{A'B'}=\tfrac{1}{2}F\fdud{CA'|}{C}{B'}$.

Another useful feature of the spinor formalism is the Fierz identity. Given three spinors, $\phi_A$, $\chi_B$, $\psi_C$, the anti-symmetrization over their indices equals zero, as their indices only run over two values $A,B,C=0,1$.  The Fierz identity is obtained by contracting this anti-symmetrized product with $\epsilon^{BC}$, which leads to
\begin{align} \label{Fierz}
    3\epsilon^{BC}\phi_{[A}\chi_{B}\psi_{C]}=\phi_{A}(\chi\psi)+\chi_{A}(\psi\phi)+\psi_{A}(\phi\chi)\equiv0 \,.
\end{align}

\section{Technicalities: SDYM}\label{app:sdym}
The calculations in the main text have been highly compacted for the sake of brevity. In this appendix we aim to present some proofs and additional details to the reader.
\subsection{\texorpdfstring{$\Psi$}{psi}-sector} \label{app:SDYMPsi}
The calculations of the $\Psi$-sector have been moved to this appendix as they are very much similar to the $F$-sector. The approach is as follows: we apply a covariant derivative to the ansatz \eqref{ansatzpsi} and we also contract $e^{BB'}$ with $\Psi_{A(k+1),A'(k+3)}$ as to obtain two expressions for $e^{BB'}D\Psi_{A(k)B,A'(k+2)B'}$, which we then compare. The former yields
\begin{align} \label{DsquaredPsi}
    \begin{aligned}
        D^2\Psi_{A(k),A'(k+2)}=&-H^{BB}[F_{BB},\Psi_{A(k),A'(k+2)}]=-e^{CC'}\wedge D\Psi_{A(k)C,A'(k+2)C'} \\
        &-e\fud{C}{A'}\wedge\sum_{n=0}^{k-1}\beta_{nk}[DF_{A(n+1)C,A'(n)},\Psi_{A(k-n-1),A'(k-n+1)}] \\
        &-e\fud{C}{A'}\wedge\sum_{n=0}^{k-1}\beta_{nk}[F_{A(n+1)C,A'(n)},D\Psi_{A(k-n-1),A'(k-n+1)}] \\
        &-e\fud{C}{A'}\wedge\sum_{n=0}^{k-1}\gamma_{nk}[DF_{A(n+2),A'(n)},\Psi_{A(k-n-2)C,A'(k-n+1)}] \\
        &-e\fud{C}{A'}\sum_{n=0}^{k-1}\gamma_{nk}[F_{A(n+2),A'(n)},D\Psi_{A(k-n-2),A'(k-n+1)}] \,.
    \end{aligned}
\end{align}
Considering only quadratic terms in the fields gives
{\allowdisplaybreaks
\begin{align} \label{nablasquaredpsi}
    \begin{aligned}
        e^{BB'}\wedge D\Psi_{A(k)C,A'(k+2)C'}&=H^{BB}[F_{BB},\Psi_{A(k),A'(k+2)}]\\
        &-\sum_{n=1}^{k}\tfrac{\beta_{(n-1)k}}{2}H^{BB}[F_{A(n)BB,A'(n)},\Psi_{A(k-n),A'(k-n+2)}]\\
        &-\sum_{n=0}^{k}\tfrac{\beta_{nk}+\gamma_{(n-1)k}}{2}H^{BB}[F_{A(n+1)B,A'(n)},\Psi_{A(k-n-1)B,A'(k-n+2)}]\\
        &-\sum_{n=0}\tfrac{\gamma_{nk}}{2}H^{BB}[F_{A(n+2),A'(n)},\Psi_{A(k-n-2)BB,A'(k-n+2)}]\\
        &+\sum_{n=0}^{k}\tfrac{\gamma_{(n-1)k}}{2}H\fdu{A'}{B'}[F\fdud{A(n+1)}{B}{,A'(n-1)B'},\Psi_{A(k-n-1)B,A'(k-n+2)}]\\
        &-\sum_{n=0}^{k}\tfrac{\beta_{nk}}{2}H\fdu{A'}{B'}[F\fdud{A(n+1)}{B}{,A'(n)},\Psi_{A(k-n-1)B,A'(k-n+1)B'}]\,.
    \end{aligned}
\end{align}}\noindent
We have renamed the dummy indices in some terms in order to match the summation limits with the expression for $e^{BB'}\Psi_{A(k)B,A'(k+2)B'}$ that we will derive next. This makes some coefficients show up that were not present in the minimal ansatz, so we have to set them to zero by hand: $\beta_{kk}=0$, $\gamma_{(-1)k}=0$. Contracting $e^{BB'}$ with $D\Psi_{A(k),A'(k+4)}$ gives
\begin{align*}
    \begin{aligned}
        &e^{BB'}\wedge D\Psi_{A(k)B,A'(k+2)B'}=\\
        &-\sum_{n=0}^{k}\tfrac{(n+1)(k+4)}{(k+1)(k+3)}\tfrac{\beta_{n(k+1)}}{2}H^{BB}[F_{A(n)BB,A'(n)},\Psi_{A(k-n),A'(k-n+2)}]\\
        &-\sum_{n=0}^{k}(\tfrac{(k-n)(k+4)}{(k+1)(k+3)}\tfrac{\beta_{n(k+1)}}{2}+\tfrac{(n+2)(k+4)}{(k+1)(k+3)}\tfrac{\gamma_{n(k+1)}}{2})H^{BB}[F_{A(n+1)B,A'(n)},\Psi_{A(k-n-1)B,A'(k-n+2)}]\\
        &-\sum_{n=0}^{k}(\tfrac{n(k-n)}{(k+1)(k+3)}\tfrac{\beta_{n(k+1)}}{2}-\tfrac{n(n+2)}{(k+1)(k+3)}\tfrac{\gamma_{n(k+1)}}{2})H\fdu{A'}{B'}[F\fdud{A(n+1)}{B}{,A'(n-1)B'},\Psi_{A(k-n-1)B,A'(k-n+2)}]\\
        &-\sum_{n=0}^{k}(\tfrac{(k-n)(k-n+2)}{(k+1)(k+3)}\tfrac{\beta_{n(k+1)}}{2}-\tfrac{(n+2)(k-n+2)}{(k+1)(k+3)}\tfrac{\gamma_{n(k+1)}}{2})H\fdu{A'}{B'}[F\fdud{A(n+1)}{B}{,A'(n)},\Psi_{A(k-n-1)B,A'(k-n+1)B'}]\\
        &-\sum_{n=0}^{k}\tfrac{(k-n-1)(k+4)}{(k+1)(k+3)}\tfrac{\gamma_{n(k+1)}}{2}H^{BB}[F_{A(n+2),A'(n)},\Psi_{A(k-n-2)BB,A'(k-n+2)}]\,.
    \end{aligned}
\end{align*}
Comparing this expression to \eqref{nablasquaredpsi}, one obtains the recurrence relations
\besubeqs
    \begin{align*}
        0&=\beta_{0k}+\tfrac{2k(k+2)}{k+3} \,, \\
        0&=\tfrac{(n+2)(k+4)}{(k+1)(k+3)}\tfrac{\beta_{(n+1)(k+1)}}{2}-\tfrac{\beta_{nk}}{2} \,, \\
        0&=\tfrac{(k-n)(k+4)}{(k+1)(k+3)}\tfrac{\beta_{n(k+1)}}{2}+\tfrac{(n+2)(k+4)}{(k+1)(k+3)}\tfrac{\gamma_{n(k+1)}}{2}-\tfrac{\beta_{nk}+\gamma_{(n-1)k}}{2} \,, \\
        0&=\tfrac{(k-n)(k-n+2)}{(k+1)(k+3)}\tfrac{\beta_{n(k+1)}}{2}-\tfrac{(n+2)(k-n+2)}{(k+1)(k+3)}\tfrac{\gamma_{n(k+1)}}{2}-\tfrac{\beta_{nk}}{2} \,, \\
        0&=\tfrac{(k-n-1)(n+1)}{(k+1)(k+3)}\tfrac{\beta_{(n+1)(k+1)}}{2}-\tfrac{(n+3)(n+1)}{(k+1)(k+3)}\tfrac{\gamma_{(n+1)(k+1)}}{2}+\tfrac{\gamma_{nk}}{2} \,, \\
        0&=\tfrac{(k-n-1)(k+4)}{(k+1)(k+3)}\tfrac{\gamma_{n(k+1)}}{2}-\tfrac{\gamma_{nk}}{2} \,.
    \end{align*}
\esubeqs
The system is solved by
\begin{align*}
    \beta_{nk}&=-\tfrac{2}{(n+1)!}\tfrac{k-n+2}{k+3}\tfrac{k!}{(k-n-1)!} \,, & \gamma_{nk}&=\tfrac{2}{(n+2)!}\tfrac{n+1}{k+3}\tfrac{k!}{(k-n-2)!} \,.
\end{align*}

\subsection{Absence of higher order corrections} \label{app:sdymTruncation}
In section \ref{sec:flatSDYM} it was mentioned that the obtained solutions for $DF_{A(k+2),A'(k)}$ and $D \Psi_{A(k),A'(k+2)}$ ensured that no higher order corrections were needed. This result is equivalent to the consistency of the $L_\infty$-relation in \eqref{SDYMstasheffF3} and \eqref{SDYMstafhessPsi3}. Here we shall present the proof.

\paragraph{$\boldsymbol{F}$-sector.} 
As a starting point we take the solution from \eqref{spin1sol} and plug it into \eqref{DsquaredF}, from which we only consider only the cubic terms. This gives us the l.h.s. of the $L_\infty$-relation \eqref{SDYMstasheffF3}:
\begin{align*}
\begin{aligned}
    &l_3(e,l_3(e,F,F),F)+l_3(e,F,l_3(e,F,F))=\\
    &-\tfrac{1}{2}H_{A'A'}\sum_{n=1}^{k-1}\sum_{m=0}^{n-1}\tfrac{n-m+1}{n+2}\alpha_{nk}\alpha_{mn}[F_{A(k-n+1),A'(k-n-1)},[F_{A(m+1)B,A'(m)},F\fdud{A(n-m)}{B}{,A'(n-m-1)}]]\\
    &+\tfrac{1}{2}H_{A'A'}\sum_{n=0}^{k-2}\sum_{m=0}^{k-n-2}\alpha_{nk}\alpha_{m(k-n-1)}[F\fdud{A(n+1)}{B}{,A'(n)},[F_{A(n+1)B,A'(m)},F_{A(k-n-m),A'(k-n-m-2)}]]\\
    &=-\tfrac{1}{2}H_{A'A'}\sum_{n=1}^{k-1}\sum_{m=0}^{n-1}\tfrac{n-m+1}{n+2}\alpha_{nk}\alpha_{mn}[F\fdud{A(m+1)}{B}{,A'(m)},[F_{A(n-m)B,A'(n-m-1)},F_{A(k-n+1),A'(k-n-1)}]]\\
    &-\tfrac{1}{2}H_{A'A'}\sum_{n=1}^{k-1}\sum_{m=0}^{n-1}\tfrac{n-m+1}{n+2}\alpha_{nk}\alpha_{mn}[F\fdud{A(n-m)}{B}{,A'(n-m-1)},[F_{A(m+1)B,A'(m)},F_{A(k-n+1),A'(k-n-1)}]]\\
    &+\tfrac{1}{2}H_{A'A'}\sum_{n=0}^{k-2}\sum_{m=0}^{k-n-2}\alpha_{nk}\alpha_{m(k-n-1)}[F\fdud{A(n+1)}{B}{,A'(n)},[F_{A(n+1)B,A'(m)},F_{A(k-n-m),A'(k-n-m-2)}]]\,,
\end{aligned}
\end{align*}
where we applied the Jacobi identity on the very first term. In order to compare the three terms on the r.h.s., the nested commutators must be cast into the same form, which can be achieved by renaming the dummy indices. The final result allows all terms to be collected into one and evaluates to
\begin{align*}
\begin{aligned}
    &\tfrac{1}{2}H_{A'A'}\sum_{n=0}^{k-2}(\sum_{m=0}^{n}\alpha_{mk}\alpha_{(n-m)(k-m-1)}-\tfrac{m+2}{n+3}\alpha_{(n+1)k}\alpha_{(n-m)(n+1)}-\tfrac{n-m+2}{n+3}\alpha_{(n+1)k}\alpha_{m(n+1)})\\
    &\times[F\fdud{A(m+1)}{B}{,A'(m)},[F_{A(n-m+1)B,A'(n-m)},F_{A(k-n),A'(k-n-2)}]]=0\,,
\end{aligned}
\end{align*}
for which the solution for $\alpha_{nk}$ was used. This proves the $L_\infty$-relation \eqref{SDYMstasheffF3}.

\paragraph{$\boldsymbol{\Psi}$-sector.}
We isolate the terms cubic in the fields in \eqref{DsquaredPsi} and we plug in \eqref{spin1sol} and \eqref{solpsi}, which yields the l.h.s. of $L_\infty$-relation \eqref{SDYMstafhessPsi3} and reads{\allowdisplaybreaks
\begin{align*}
    \begin{aligned}
        &l_3(e,l_3(e,F,F),\Psi)+l_3(e,F,l_3(e,F,\Psi))=\\
        &H_{A'A'}\sum_{n=1}^{k-1}\sum_{m=0}^{n-1}\tfrac{\alpha_{mn}\beta_{nk}}{2}\tfrac{n-m+1}{n+2}[[F_{A(m+1)B,A'(m)},F\fdud{A(n-m)}{B}{,A'(n-m-1)}],\Psi_{A(k-n-1),A'(k-n+1)}]\\
        &+H_{A'A'}\sum_{n=0}^{k-2}\sum_{m=0}^{k-n-2}\tfrac{\beta_{m(k-n-1)}\beta_{nk}}{2}[F\fdud{(n+1)}{B}{,A'(n)},[F_{A(m+1)B,A'(m)},\Psi_{A(k-n-m-2),A'(k-n-m))}]]\\
        &+H_{A'A'}\sum_{n=0}^{k-3}\sum_{m=0}^{k-n-3}\tfrac{\beta_{nk}\gamma_{m(k-n-1)}}{2}[F\fdud{A(n+1)}{B}{,A'(n)},[F_{A(m+2),A'(m)},\Psi_{A(k-n-m-3)B,A'(k-n-m)}]]\\
        &+H_{A'A'}\sum_{n=1}^{k-2}\sum_{m=0}^{n-1}\tfrac{\alpha_{mn}\gamma_{nk}}{2}[[F_{A(m+1)B,A'(m)},F_{A(n-m+1),A'(n-m-1)}],\Psi\fdud{A(k-n-2)}{B}{,A'(k-n+1)}]\\
        &+H_{A'A'}\sum_{n=0}^{k-2}\sum_{m=0}^{k-n-3}(\tfrac{\beta_{m(k-n-1)\gamma_{nk}}}{2}\tfrac{k-n-m-2}{k-n-1}-\tfrac{\gamma_{m(k-n-1)\gamma_{nk}}}{2}\tfrac{m+2}{k-n-1})\\
        &     \times[F_{A(n+2),A'(n)},[F\fdud{A(m+1)}{B}{,A'(m)},\Psi_{A(k-n-m-3)B,A'(k-n-m)}]].
    \end{aligned}
\end{align*}}\noindent
Our approach is similar to the one for the $F$-sector: we aim to reduce the equations as much as possible by casting the nested commutators into a similar form. A particular technicality in this case is that a contraction can be either between two $F$'s or between $F$ and $\Psi$. The Fierz identity is used to convert all contractions into the latter type. However, one must be careful, as the Fierz identity requires some free indices on the available spinors, which might not be present in all terms of the summation. Hence we isolate these cases and check that their contribution vanishes.
\begin{align*}
    \begin{aligned}
        &H_{A'A'}\sum_{n=0}^{k-2}\tfrac{\beta_{(k-1)k}\alpha_{n(k-1)}}{2}\tfrac{k-n}{k+1}[[F_{A(n+1)B,A'(n)},F\fdud{A(k-n-1)}{B}{,A'(k-n-2)}],\Psi_{A'A'}]\\
        &+H_{A'A'}\sum_{n=0}^{k-2}\tfrac{\beta_{nk}\beta_{(k-n-2)(k-n-1)}}{2}[F\fdud{A(n+1)}{B}{,A'(n)},[F_{A(k-n-1)B,A'(k-n-2)},\Psi_{A'A'}]]\\
        &=H_{A'A'}\sum_{n=0}^{k-2}(-\tfrac{\beta_{(k-1)k}\alpha_{n(k-1)}}{2}\tfrac{k-n}{k+1}+\tfrac{\beta_{nk}\beta_{(k-n-2)(k-n-1)}}{2}-\tfrac{\beta_{(k-1)k}\alpha_{(k-n-2)(k-1)}}{2}\tfrac{n+2}{k+1})\\
        &   \times[F\fdud{A(n+1)}{B}{,A'(n)},[F_{A(k-n-1)B,A'(k-n-2)},\Psi_{A'A'}]]=0\,,
    \end{aligned}
\end{align*}
where we used the solution for $\beta_{nk}$ and $\alpha_{nk}$. Finally, applying the Fierz identity, Jacobi identity and renaming of dummy indices allows one to cast the remaining terms into a more practical form that reads
\begin{align*}
    \begin{aligned}
        &H_{A'A'}\sum_{ n=0}^{k-3}\sum_{m=0}^{n}(\tfrac{\beta_{(n+1)k}\alpha_{m(n+1)}}{2}\tfrac{n-m+2}{n+3}+\tfrac{\beta_{(n+1)k}\alpha_{(n-m)(n+1)}}{2}\tfrac{m+2}{n+3}+\tfrac{\beta_{mk}\beta_{(n-m)(k-m-1)}}{2}\\
        &+\tfrac{\beta_{mk}\gamma_{(n-m)(k-m-1)}}{2}-\tfrac{\gamma_{(n+1)k}\alpha_{m(n+1)}}{2})[F\fdud{A(m+1)}{B}{,A'(m)},[F_{A(n-m+2),A'(n-m)},\Psi_{A(k-n-3)B,A'(k-n)}]]\\
        &+H_{A'A'}\sum_{n=0}^{k-3}\sum_{m=0}^{n}(\tfrac{\beta_{(n+1)k}\alpha_{(n-m)(n+1)}}{2}\tfrac{m+2}{n+3}+\tfrac{\beta_{(n+1)k}\alpha_{m(n+1)}}{2}\tfrac{n-m+2}{n+3}-\tfrac{\beta_{mk}\beta_{(n-m)(k-m-1)}}{2}\\
        &+\tfrac{\gamma_{(n+1)k}\alpha_{(n-m)(n+1)}}{2}-\tfrac{\gamma_{mk}\beta_{(n-m)(k-m-1)}}{2}\tfrac{k-n-2}{k-m-1}+\tfrac{\gamma_{mk}\gamma_{(n-m)(k-m-1)}}{2}\tfrac{n-m+2}{k-m-1})\\
        &\times[F_{A(m+2),A'(m)},[F\fdud{A(n-m+1)}{B}{,A'(n-m)},\Psi_{A(k-n-3)B,A'(k-n)}]]=0 \,,
    \end{aligned}
\end{align*}
which is obtained by plugging in the solutions for $\alpha_{nk}$, $\beta_{nk}$ and $\gamma_{nk}$ were used. This implies the consistency of $L_\infty$-relation \eqref{SDYMstafhessPsi3}.

\subsection{Higher gravitational corrections} \label{app:sdymLambda}
In section \ref{sec:SDYMcurved} we mentioned that the correction due to the constant gravitational background to the linear term in $D F_{A(k+2),A'(k)}$ and $D\Psi_{A(k),A'(k+2)}$ does not propagate to the quadratic term or higher. This appendix is dedicated to prove this.

\paragraph{$\boldsymbol{F}$-sector.}
The $L_\infty$-relations are modified on a constant curvature background according to \eqref{SDYMstasheffLambda}. It was mentioned in \eqref{SDYMstasheffLambda} that the gravitational contribution decouples and vanishes independently. We shall present a proof here.

We are interested in checking consistency of $L_\infty$-relation \eqref{stasheffLambdaF}. We do so by taking the covariant derivative of \eqref{curvedFsol}, which gives
\begin{align} \label{DsquaredFLambda}
    \begin{aligned}
        D^2F_{A(k+2),A'(k)}&=-H^{BB}[F_{BB},F_{A(k+2),A'(k)}]+(k+2)H\fdu{A}{B}F_{A(k+1)B,A'(k)}\\
        &+k H\fdu{A'}{B'}F_{A(k+2),A'(k-1)B'}=-e^{BB'}\wedge DF_{A(k+2)B,A'(k)B'}\\
        &-e\fud{B}{A'}\wedge\sum_{n=0}^{k-1}\alpha_{nk}[DF_{A(n+1)B,A'(n)},F_{A(k-n+1),A'(k-n-1)}] \\
        &-e\fud{B}{A'}\wedge\sum_{n=0}^{k-1}\alpha_{nk}[F_{A(n+1)B,A'(n)},DF_{A(k-n+1),A'(k-n-1)}] \,.
    \end{aligned}
\end{align}
Considering only the terms coming from gravitational contributions gives the l.h.s of $L_\infty$-relation \eqref{stasheffLambdaF} and reads after introducing $f_{k}=k(k+2)$:
\begin{align*}
    \begin{aligned}
        &\tilde{l}_2(e,l_3(e,F,F))+l_3(e,\tilde{l}_2(e,F),F)+l_3(e,F,\tilde{l}_2(e,F))=\\
        &H_{A'A'}\sum_{n=0}^{k-2}\tfrac{\tfrac{1}{2}n+2}{n+3}f_{n+1}\alpha_{(n+1)k}[F_{A(n+2),A'(n)},F_{A(k-n),A'(k-n-2)}]\\
        &+H_{A'A'}\sum_{n=0}^{k-2}\tfrac{\alpha_{nk}}{2}f_{k-n-1}[F_{A(n+2),A'(n)},F_{A(k-n),A'(k-n-2)}]\\
        &-H_{A'A'}\sum_{n=0}^{k-2}\tfrac{\alpha_{n(k-1)}}{2}f_k[F_{A(n+2),A'(n)},F_{A(k-n),A'(k-n-2)}]\\
        &= \sum_{n=0}^{k-2}\tfrac{1}{2}(\tfrac{\tfrac{1}{2}n+2}{n+3}f_{n+1}\alpha_{(n+1)k}-\tfrac{\tfrac{1}{2}(k-n)+1}{k-n+1}f_{k-n-1}\alpha_{(k-n-1)k}+\tfrac{\alpha_{nk}}{2}f_{k-n-1}-\tfrac{\alpha_{(k-n-2)k}}{2}f_{n+1}\\
        &-\tfrac{\alpha_{n(k-1)}}{2}f_k+\tfrac{\alpha_{(k-n-2)(k-1)}}{2}f_k)[F_{A(n+2),A'(n)},F_{A(k-n),A'(k-n-2)}]=0\,,
    \end{aligned}
\end{align*}
where the anti-symmetry of the commutator has been made explicit and the solution for $\alpha_{nk}$ was applied. Thus, the modification to the second $L_\infty$-relation of the $F$-sector vanishes, which means that the gravitational background only modifies $DF_{A(k+2),A'(k)}$ on the linear level, identically to the free equations. This is equivalent to the consistency of \eqref{stasheffLambdaF}.

\paragraph{$\boldsymbol{\Psi}$-sector.}
The second $L_\infty$-relation for $\Psi$ on a gravitational background is modified according to \eqref{stasheffLambdaPsi}. This gives
\begin{align} \label{DsquaredPsiLambda}
    \begin{aligned}
        D^2\Psi_{A(k),A'(k+2)}=&-H^{BB}[F_{BB},\Psi_{A(k),A'(k+2)}]+kH\fdu{A}{B}\Psi_{A(k-1)B,A'(k+2)}\\
        &+(k+2)H\fdu{A'}{B'}\Psi_{A(k),A'(k+1)B'}=-e^{CC'}\wedge D\Psi_{A(k)C,A'(k+2)C'} \\
        &-e\fud{C}{A'}\wedge\sum_{n=0}^{k-1}\beta_{nk}[DF_{A(n+1)C,A'(n)},\Psi_{A(k-n-1),A'(k-n+1)}] \\
        &-e\fud{C}{A'}\wedge\sum_{n=0}^{k-1}\beta_{nk}[F_{A(n+1)C,A'(n)},D\Psi_{A(k-n-1),A'(k-n+1)}] \\
        &-e\fud{C}{A'}\wedge\sum_{n=0}^{k-1}\gamma_{nk}[DF_{A(n+2),A'(n)},\Psi_{A(k-n-2)C,A'(k-n+1)}] \\
        &-e\fud{C}{A'}\sum_{n=0}^{k-1}\gamma_{nk}[F_{A(n+2),A'(n)},D\Psi_{A(k-n-2),A'(k-n+1)}] \,.
    \end{aligned}
\end{align}
Considering only the terms containing a gravitational contribution, one obtains the l.h.s. of the $L_\infty$-relation \eqref{stasheffLambdaPsi}:
\begin{align}
    \begin{aligned}
         &\tilde{l}_2(e,l_3(e,F,\Psi))+l_3(e,\tilde{l}_2(e,F),\Psi)+l_3(e,F,\tilde{l}_2(e,\Psi))\\
         &=H_{A'A'}\sum_{n=0}^{k-2}(f_{n+1}\beta_{(n+1)k}\tfrac{\tfrac{1}{2}n+2}{n+3}+\tfrac{1}{2}\beta_{nk}f_{k-n-1}+\tfrac{1}{2}f_{n+1}\gamma_{(n+1)k}\\
        &+\tfrac{\tfrac{1}{2}k-\tfrac{1}{2}n}{k-n-1}\gamma_{nk}f_{k-n-1}-f_k\tfrac{\beta_{n(k-1)}}{2}-f_k\tfrac{\gamma_{n(k-1)}}{2})[F_{A(n+2),A'(n)},\Psi_{A(k-n-2),A'(k-n)}]=0\,,
    \end{aligned}
\end{align}
which we obtain by plugging in the solutions for $\alpha_{nk}$, $\beta_{nk}$ and $\gamma_{nk}$. This proves the consistency of \eqref{stasheffLambdaPsi}.

The results in this appendix prove that the gravitational contribution to the $L_\infty$-relations in both sectors decouples and vanishes independently, which is equivalent to consistency of \eqref{stasheffLambdaF} and \eqref{stasheffLambdaPsi}. Thus, the gravitational background only modifies
$DF_{A(k+2),A'(k)}$ and $D\Psi_{A(k),A'(k+2)}$ on the linear level, identically to the free equations.

\section{Technicalities: SDGR}
Several technicalities have been left out from the main text. In this section we aim to present the calculation of the $\Psi$-sector, as well as the proofs of the truncation of $\nabla C$ and $\nabla\Psi$, as promised in section \ref{sec:SDGRflat}
\subsection{\texorpdfstring{$\Psi$}{psi}-sector} \label{app:SDGRPsi}
We have left the details of the calculation of the $\Psi$-sector of section \ref{sec:SDGRflat} to this appendix, as it bears a lot of resemblance to the $C$-sector.

The approach is similar to before: we take the covariant derivative of the ansatz \eqref{eq:SDGRPsiansatz} and we also contract $e^{BB'}$ with $\nabla\Psi_{A(k+1),A'(k+5)}$ as this will give two expressions for $e^{BB'}\wedge\nabla\Psi_{A(k)B,A'(k+4)B'}$, so we can compare them. This will unveil its structure. The former yields{\allowdisplaybreaks
\begin{align}\label{nablasquaredPsi}
    \begin{aligned}
        \nabla^2\Psi_{A(k),A'(k+4)}&=kH^{BB}C\fdu{ABB}{D}\Psi_{A(k-1)D,A'(k+4)}=-e^{CC'}\wedge\nabla\Psi_{A(k)C,A'(k+4)C'}\\
        &-e\fud{C}{A'}\wedge\sum_{n=0}^{k}b_{nk}\nabla C\fdud{A(n+2)C}{D}{,A'(n)}\Psi_{A(k-n-2)D,A'(k-n+3)}\\
        &-e\fud{C}{A'}\wedge\sum_{n=0}^{k}b_{nk} C\fdud{A(n+2)C}{D}{,A'(n)}\nabla\Psi_{A(k-n-2)D,A'(k-n+3)}\\
        &-e\fud{C}{A'}\wedge\sum_{n=0}^{k}c_{nk}\nabla C\fdud{A(n+3)}{D}{,A'(n)}\Psi_{A(k-n-3)CD,A'(k-n+3)}\\
        &-e\fud{C}{A'}\wedge\sum_{n=0}^{k}c_{nk} C\fdud{A(n+3)}{D}{,A'(n)}\nabla\Psi_{A(k-n-3)CD,A'(k-n+3)} \,.
    \end{aligned}
\end{align}}\noindent
Isolating the terms quadratic in the fields gives
\begin{align*}
    \begin{aligned}
        e^{BB'}\wedge\nabla\Psi_{A(k)C,A'(k+4)C'}&= -kH^{BB}C\fdu{ABB}{D}\Psi_{A(k-1)D,A'(k+4)}\\
        &-\frac{1}{2}H^{BB}\sum_{n=0}^{k}b_{nk}C\fdud{A(n+2)BB}{D}{,A'(n+1)}\Psi_{A(k-n-2)D,A'(k-n+3)}\\
        &-\frac{1}{2}H^{BB}\sum_{n=0}^{k}(b_{nk}+c_{(n-1)k})C\fdud{A(n+2)B}{D}{,A'(n)}\Psi_{A(k-n-2)BD,A'(k-n+4)}\\
        &+\frac{1}{2}H\fdu{A'}{B'}\sum_{n=0}^{k}b_{nk}C\fdud{A(n+2)}{BD}{,A'(n)}\Psi_{A(k-n-2)BD,A'(k-n+3)B'}\\
        &-\frac{1}{2}H\fdu{A'}{B'}\sum_{n=0}^{k}c_{nk}C\fdud{A(n+3)}{BD}{,A'(n)B'}\Psi_{A(k-n-3)BD,A'(k-n+3)}\\
        &-\frac{1}{2}H^{BB}\sum_{n=0}^{k}c_{nk}C\fdud{A(n+3)}{D}{,A'(n)}\Psi_{A(k-n-3)BBD,A'(k-n+4)} \,,
    \end{aligned}
\end{align*}
whereas the latter gives
{\allowdisplaybreaks
\begin{align*}
        &e^{BB'}\wedge\nabla\Psi_{A(k)B,A'(k+4)B'}=-H^{BB}b_{0(k+1)}\tfrac{k+6}{(k+1)(k+5)}C\fdu{ABB}{D}\Psi_{A(k-1)D,A'(k+4)}\\
        &-\frac{1}{2}H^{BB}\sum_{n=0}^kb_{(n+1)(k+1)}\tfrac{(k+6)(n+3)}{(k+1)(k+5)}C\fdud{A(n+2)BB}{D}{,A'(n+1)}\Psi_{A(k-n-2)D,A'(k-n+3)}\\
        &-\tfrac{1}{2}H^{BB}\sum_{n=0}^{k}(b_{n(k+1)}\tfrac{(k+6)(k-n-1)}{(k+1)(k+5)}+c_{n(k+1)}\tfrac{(k+6)(n+3)}{(k+1)(k+5)})\\
        &\times C\fdud{A(n+2)B}{D}{,A'(n)}\Psi_{A(k-n-2)BD,A'(k-n+4)}\\
        &+\frac{1}{2}H\fdu{A'}{B'}\sum_{n=0}^{k}(b_{(n+1)(k+1)}\tfrac{(n+1)(k-n-2)}{(k+1)(k+5)}-c_{(n+1)(k+1)}\tfrac{(n+1)(n+4)}{(k+1)(k+5)})\\
        &\times C\fdud{A(n+3)}{BD}{,A'(n)B'}\Psi_{A(k-n-3)BD,A'(k-n+3)}\\
        &+\frac{1}{2}H\fdu{A'}{B'}\sum_{n=0}^{k}(b_{n(k+1)}\tfrac{(k-n+4)(k-n-1)}{(k+1)(k+5)}-c_{n(k+1)}\tfrac{(k-n+4)(n+3)}{(k+1)(k+5)})\\
        &\times C\fdud{A(n+2)}{BD}{,A'(n)}\Psi_{A(k-n-2)BD,A'(k-n+3)B'}\\
        &-\frac{1}{2}H^{BB}\sum_{n=0}^{k}c_{n(k+1)}\tfrac{(k+6)(k-n-2)}{(k+1)(k+5)}C\fdud{A(n+3)}{D}{,A'(n)}\Psi_{A(k-n-3)BBD,A'(k-n+4)} \,.
\end{align*}}\noindent
Comparing them gives the system of recurrence relations
{\allowdisplaybreaks
\besubeqs
    \begin{align*}
        0&=\tfrac{k+6}{(k+1)(k+5)}b_{0(k+1)}-k\,,\\
        0&=b_{nk}-b_{(n+1)(k+1)}\tfrac{(k+6)(n+3)}{(k+1)(k+5)}\,,\\
        0&=b_{nk}+c_{nk}-b_{n(k+1)}\tfrac{(k+6)(k-n-1)}{(k+1)(k+5)}-c_{n(k+1)}\tfrac{(k+6)(n+3)}{(k+1)(k+5)}\,,\\
        0&=b_{nk}-b_{n(k+1)}\tfrac{(k-n+4)(k-n-1)}{(k+1)(k+5)}+c_{n(k+1)}\tfrac{(k-n+4)(n+3)}{(k+1)(k+5)}\,,\\
        0&=c_{nk}+b_{(n+1)(k+1)}\tfrac{(n+1)(k-n-2)}{(k+1)(k+5)}-c_{(n+1)(k+1)}\tfrac{(n+1)(n+4)}{(k+1)(k+5)}\,,\\
        0&=c_{nk}-c_{n(k+1)}\tfrac{(k+6)(k-n-2)}{(k+1)(k+5)}\,,
    \end{align*}
\esubeqs}\noindent
which is solved by
    \begin{align*}
        b_{nk}&=\tfrac{2}{(n+2)!}\tfrac{k!}{(k-n-2)!}\tfrac{k-n+4}{k+5}&
        c_{nk}&=-\tfrac{2}{(n+2)!}\tfrac{k!}{(k-n-3)!}\tfrac{n+1}{(k+5)(n+3)}\,.
    \end{align*}
\subsection{Absense of higher order corrections} \label{app:sdgrTruncation}
\paragraph{$\boldsymbol{C}$-sector.} We consider \eqref{nablasquaredC} and isolate the terms cubic in $C$. Plugging in the solution from \eqref{Csol} yields the l.h.s. of the $L_\infty$-relation \eqref{stasheffC3} given by
{\allowdisplaybreaks
\begin{align} \label{Ccubed}
    \begin{aligned}
        &l_3(e,l_3(e,C,C),C)+l_3(e,C,l_3(e,C,C))=\\
        &\tfrac{1}{2}H_{A'A'}\sum_{n=1}^{k-1}\sum_{m=0}^{n-1}a_{nk}a_{mn}\tfrac{n-m+2}{n+4}\tfrac{m+2}{n+3} C\fdud{A(m+1)B}{DE}{,A'(m)}C\fdud{A(n-m+1)E}{B}{,A'(n-m-1)}C_{A(k-n+2)D,A'(k-n-1)}\\
        &+\tfrac{1}{2}H_{A'A'}\sum_{n=1}^{k-1}\sum_{m=0}^{n-1}a_{nk}a_{mn}\tfrac{n-m+2}{n+4}\tfrac{n-m+1}{n+3}\\
        & \times C\fdud{A(m+2)B}{E}{,A'(m)}C\fdud{A(n-m)E}{BD}{,A'(n-m-1)}C_{A(k-n+2)D,A'(k-n-1)}\\
        &+\tfrac{1}{2}H_{A'A'}\sum_{n=0}^{k-2}\sum_{m=0}^{k-n-2}a_{nk}a_{m(k-n-1)}\tfrac{m+2}{k-n+3}\\
        &  C\fdud{A(n+2)}{BD}{,A'(n)}C\fdud{A(m+1)BD}{E}{,A'(m)}C_{A(k-n-m+1)E,A'(k-n-m-2)}\\
        &+\tfrac{1}{2}H_{A'A'}\sum_{n=0}^{k-2}\sum_{m=0}^{k-n-2}a_{nk}a_{m(k-n-1)}\tfrac{k-n-m+1}{k-n+3}\\
        &\times C\fdud{A(n+2)}{BD}{,A'(n)}C\fdud{A(m+2)B}{E}{,A'(m)}C_{A(k-n-m)DE,A'(k-n-m-2)}\,.
    \end{aligned}
\end{align}}\noindent
The first three terms can be collected into
\begin{align*}
    \begin{aligned}
        &\tfrac{1}{2}H_{A'A'}\sum_{n=1}^{k-1}\sum_{m=0}^{n-1}(a_{nk}a_{mn}\tfrac{n-m+2}{n+4}\tfrac{m+2}{n+3}+a_{nk}a_{(n-m-1)n}\tfrac{m+3}{n+4}\tfrac{m+2}{n+3}-a_{(n-m-1)k}a_{m(k-n+m)}\tfrac{m+2}{k-n+m+4})\\
        &\times C\fdud{A(m+1)B}{DE}{,A'(m)}C\fdud{A(n-m+1)}{B}{,A'(n-m-1)}C_{A(k-n+2)D,A'(k-n-1)}=0\,,
    \end{aligned}
\end{align*}
for which the solution for $a_{nk}$ is applied. The last term in \eqref{Ccubed} may be rewritten as
\allowdisplaybreaks{
\begin{align*}
    \begin{aligned}
        &\tfrac{1}{2}H_{A'A'}\sum_{n=0}^{k-2}\sum_{m=0}^{n}a_{mk}a_{(n-m)(k-m-1)}\tfrac{k-n+1}{k-m+3} C\fdud{A(m+2)}{BD}{,A'(m)}C\fdud{A(n-m+2)B}{E}{,A'(m)}C_{A(k-n)DE,A'(k-n-2)}\\
        &=\tfrac{1}{4}H_{A'A'}\sum_{n=0}^{k-2}\sum_{m=0}^{n}(a_{mk}a_{(n-m)(k-m-1)}\tfrac{k-n+1}{k-m+3}-a_{(n-m)k}a_{m(k-n+m-1)}\tfrac{k-n+1}{k-n+m+3})\\
        &\times C\fdud{A(m+2)}{BD}{,A'(m)}C\fdud{A(n-m+2)B}{E}{,A'(m)}C_{A(k-n)DE,A'(k-n-2)}=0\,,
    \end{aligned}
\end{align*}}\noindent
where again we used the solution for $a_{nk}$. This proves that the $C$-sector truncates at quadratic order. This confirms the consistency of \eqref{stasheffC3}.

\paragraph{$\boldsymbol{\Psi}$-sector.}
We consider the cubic terms in \eqref{nablasquaredPsi} and we assume the solutions \eqref{Csol} and \eqref{psisolSDGR}. This gives the l.h.s. of the $L_\infty$-relation in \eqref{stasheffSDGRPsi3} and reads
{\allowdisplaybreaks
\begin{align*}
        &l_3(e,l_3(e,C,C),\Psi)+l_3(e,C,l_3(e,C,\Psi))=\\
        &\tfrac{1}{2}H_{A'A'}\sum_{n=0}^{k-3}\sum_{m=0}^{n}(a_{m(n+1)}b_{(n+1)k}\tfrac{(n-m+3)(m+2)}{(n+5)(n+4)}+a_{(n-m)(n+1)}b_{(n+1)k}\tfrac{(m+3)(m+2)}{(n+5)(n+4)}\\
        &-b_{m(k-n+m-1)}b_{(n-m)k}\tfrac{m+2}{k-n+m-1})\\
        &\times C\fdud{A(m+1)B}{DE}{,A'(m)}C\fdud{A(n-m+2)E}{B}{,A'(n-m)}\Psi_{A(k-n-3)D,A'(k-n+2)}\\
        &+\tfrac{1}{4}H_{A'A'}\sum_{n=0}^{k-4}\sum_{m=0}^{n}(b_{(n-m)(k-m-1)}b_{mk}\tfrac{k-n-3}{k-m-1}+c_{(n-m)(k-m-1)}b_{mk}\tfrac{n-m+3}{k-m-1}\\
        &-a_{m(n+1)}c_{(n+1)k}\tfrac{n-m+3}{n+5}-b_{m(k-n+m-1)}b_{(n-m)k}\tfrac{k-n-3}{k-n+m-1}\\
        &-c_{m(k-n+m-1)}b_{(n-m)k}\tfrac{m+3}{k-n+m-1}+a_{(n-m)(n+1)}c_{(n+1)k}\tfrac{m+3}{n+5})\\
        &\times C\fdud{A(m+2)}{BD}{,A'(m)}C\fdud{A(n-m+2)B}{E}{,A'(n-m)}\Psi_{A(k-n-4)DE,A'(k-n+2)}\\
        &+\tfrac{1}{2}H_{A'A'}\sum_{n=0}^{k-4}\sum_{m=n}^{n}(b_{(n-m)(k-m-1)}c_{mk}\tfrac{(k-n-3)(n-m+2)}{(k-m-1)(k-m-2)}\\
        &-c_{(n-m)(k-m-1)}c_{mk}\tfrac{(n-m+3)(n-m+2)}{(k-m-1)(k-m-2)}-a_{(n-m)(n+1)}c_{(n+1)k}\tfrac{n-m+2}{n+5})\\
        &\times C\fdud{A(m+3)}{D}{,A'(m)}C\fdud{A(n-m+1)BD}{E}{,A'(n-m)}\Psi\fdud{A(k-n-4)E}{B}{,A'(k-n+2)}\\
        &+\tfrac{1}{2}H_{A'A'}\sum_{n=0}^{k-5}\sum_{m=0}^{n}(c_{m(k-n+m-1)}b_{(n-m)k}\tfrac{k-n-4}{k-n+m-1}\\
        &-b_{(n-m)(k-m-1)}c_{mk}\tfrac{(k-n-3)(k-n-4)}{(k-m-1)(k-m-2)}+c_{(n-m)(k-m-1)}c_{mk}\tfrac{(n-m+3)(k-n-4)}{(k-m-1)(k-m-2)})\\
        &C\fdud{A(m+3)}{C}{,A'(m)}C\fdud{A(n-m+2)}{BD}{,A'(n-m)}\Psi_{A(k-n-5)BDE,A'(k-n+2)}=0\,,
\end{align*}}\noindent
which is obtained by plugging in the results for $a_{nk}$, $b_{nk}$ and $c_{nk}$. This proves consistency of \eqref{stasheffSDGRPsi3}.

\footnotesize
\providecommand{\href}[2]{#2}\begingroup\raggedright\endgroup

\end{document}